# Atomic-scale Imaging of Time Reversal Breaking Supercurrents in $Bi_2Sr_2CaCu_2O_8$

Bin Hu[1,2†], Catherine Dawson[1†], Shuqiu Wang[3], Siyuan Wan[1], Huiyu Zhao[1], Kuanysh Zhussubekov[1,4], S. Ishida[5], Hiroshi Eisaki[5], Peter J. Hirschfeld[2,6], Jonathan Ward[1] and J.C. Séamus Davis[1,2,7]

[1] *School of Physics, University College Cork, Cork, T12 R5C, Ireland.*
[2] *Clarendon Laboratory, Oxford University, Oxford, OX1 3PU, UK*
[3] *H. H. Wills Physics Laboratory, University of Bristol, Bristol, BS8 1TL, UK*
[4] *School of Physics, Trinity College Dublin, Dublin, D02 PN40, Ireland*
[5] *Inst. of Advanced Industrial Science and Tech., Tsukuba, Ibaraki 305-8568, Japan.*
[6] *Department of Physics, University of Florida, Gainesville FL, USA.*
[7] *Max-Planck Institute for Chemical Physics of Solids, D-01187 Dresden, Germany.*

ABSTRACT. A persistent mystery of $CuO_2$-based high temperature superconductivity[1] has been the mechanism and consequences of time reversal ($\mathcal{T}$) symmetry breaking[2-11]. Although theoretical hypotheses[12-23] focus upon persistent electronic currents as a potential cause, these have been invisible to direct experimental probes. That impasse has recently been removed by development of techniques for atomic-scale supercurrent visualization[24]. Here, by imaging the energy-resolved density of electronic states $N(r,E)$ in $Bi_2Sr_2CaCu_2O_8$, we search for the specific signatures of flowing supercurrents in a disordered $d-$wave superconductor. Employing the strategy of artificial neural networks (ANN) for complex quantum matter visualization[25], we designed a dual-ANN (VMAP) to efficiently implement $d-$wave supercurrent visualization[24], achieving imaging of supercurrent speeds $V(r)$ at magnetic fields $|B| \leq 50\,\mu T$ in $Bi_2Sr_2CaCu_2O_8$. Correspondence between specific characteristics of $N(r,E)$ around each supercurrent loop and a quantum "Rosetta Stone" which translates $N(r,E)$ to $V(r)$, emphasizes the validity of $V(r)$ detection and visualization. Then, for hole-densities $0.17 \leq p \leq 0.22$ we observe statistically repeatable networks of supercurrent flows. Crucially, these supercurrents satisfy Kirchhoff's law at junctions within each network. Further, theoretical predictions for supercurrent speeds $V$ in disordered $d-$wave superconductors[22] are consistent with measured $1250\ \mathrm{m/s} < V < 4000\ \mathrm{m/s}$, and with reported magnetic field amplitudes $|B| \approx 0.1$ mT Thus, detection and visualization of these supercurrent loop networks provides unique perspectives on $\mathcal{T}$ −symmetry breaking in cuprates. It also emphasizes the accelerating utility of ANN techniques for physics discovery from high data volume quantum matter research[25].

Hole doping antiferromagnetic insulating cuprates, converts the $CuO_2$-plane into an intensely correlated metal[26,27] whose antiferromagnetic spin-spin interactions apparently subtend high-temperature superconductivity. Yet its Hamiltonian sustains many possible orders, including $\mathcal{T}-$symmetry breaking states[28]. Cuprate superconductivity exhibits an order-parameter $\Delta(\boldsymbol{r}) = \Delta_d e^{i\theta(\boldsymbol{r})}$, where $\theta(\boldsymbol{r})$ is the macroscopic quantum phase and $\Delta_d$ exhibits $d_{x^2-y^2}$ symmetry. In these materials, $\mathcal{T}$ −symmetry breaking has long been reported. For example, zero-field muon spin relaxation ($\mu SR$) and neutron diffraction experiments in $YBa_2Cu_3O_{6+x}$ report spontaneous static magnetic fields[2,5] and in $Bi_2Sr_2CaCu_2O_8$ quasistatic magnetic fields[3], into the superconducting phase. Circularly polarized angle resolved photoemission (ARPES) studies, detect spontaneous dichroism in $Bi_2Sr_2CaCu_2O_8$ implying underlying $\mathcal{T}-$symmetry breaking[4]. Neutron scattering reveals intra-unit-cell magnetic order in $YBa_2Cu_3O_{6+x}$ and $HgBa_2CuO_4$, hypothesized due to $\mathcal{T}-$symmetry breaking loop currents[6,7] into the superconducting phase. Polar Kerr rotation in $YBa_2Cu_3O_{6+x}$ represents a non-reciprocity forbidden without $\mathcal{T}$ −symmetry breaking[8]. Second harmonic generation and birefringence also indicate breaking of $\mathcal{T}$ −symmetry[9]. Finally, the Josephson diode effect in $Bi_2Sr_2CaCu_2O_8$ pristine/twist junctions requires $\mathcal{T}-$ symmetry breaking superconductivity[10,11]. Because the long-range magnetic orders are absent in these systems, persistent atomic-scale supercurrents are hypothesized[12] as the cause of $\mathcal{T}-$symmetry breaking. But no direct evidence of such supercurrents has ever been identified.

Theoretical hypotheses for the mechanism of cuprate $\mathcal{T}-$symmetry breaking are legion[12-23]. Crystalline disorder, at surfaces or grain boundaries[14,15,23], is predicted to break $\mathcal{T}-$symmetry. However, such crystal defects are not observed in our studies. Another possibility is spatially homogeneous $\mathcal{T}$ −symmetry breaking due to a mixed singlet-triplet pairing state with $d + ip$ symmetry[16]. Alternatively, there are predictions that intra-unit-cell persistent loop currents in the pseudogap phase, not only break $\mathcal{T}$ −symmetry[17,18], but also generate currents flowing along the edges of domains of differing intra-unit-cell ordering[18]. Figure 1a is a schematic of current trajectories within a disordered intra-unit-cell loop-current system[18]. More recently, studies of highly disordered $d-$wave superconductors predict $\mathcal{T}-$ symmetry breaking[19,20,21,22]. Here, misorientation of the quantum phase difference between adjacent superconductive domains generates supercurrent loops due to random-in-sign Josephson couplings[21] (Fig. 1b). Alternatively, concentrations of intense scattering by impurity or dopant atoms may generate $d + is$ superconductivity with $\Delta_d e^{i\phi_d} + \Delta_s e^{i\phi_s}$ symmetry surrounding impurity sites, and thus supercurrents breaking $\mathcal{T}$ −symmetry[22,23] (Fig. 1c). Evaluation by direct visualization of any of these concepts has been thwarted by the absence of techniques for atomic scale visualization of supercurrents.

Recently, however, scanning tunneling microscopy techniques for atomic-scale supercurrent visualization[24] have been developed. Here, the energy spectrum $E_{\boldsymbol{k}}$ of Bogoliubov quasiparticle states $|\boldsymbol{k}\rangle$ in a superconductor is $E_{\boldsymbol{k}} = \pm\sqrt{E^2(k) + \Delta_S^2}$, where $E(k)$ is the normal-state band structure referenced to the Fermi energy and $\Delta_S$ is the electron-pairing order parameter. In a laboratory/crystal frame relative to which supercurrent flows at velocity $\boldsymbol{V}(\boldsymbol{r}) \neq 0$, each $E_{\boldsymbol{k}}$ receives a Galilean energy-boost[24, 29, 30, 31] yielding a new spectrum:

$$E_{\boldsymbol{k}}' = \pm\sqrt{E^2(k) + \Delta_S^2} + \hbar\boldsymbol{k}\cdot\boldsymbol{V}(\boldsymbol{r}) \qquad (1)$$

Figure 1d represents schematic consequences of Eqn. 1 for a $d$-wave superconductor. Importantly, the specifics of each consequent $N(E)$ spectrum depend on the vectorial direction of the supercurrent flow. Considering a $d_{x^2-y^2}$ symmetry superconductor (Fig. 1e) the resultant $N(E)$ spectra for supercurrents at angles $0, \pi/8, \pi/4$ and $3\pi/8$ relative to the order parameter and for different supercurrent speeds $\boldsymbol{V}$, are shown in Figs. 1f-i. This is a quantum "Rosetta Stone" translating theoretic spectral shapes $N(r,E)$ into supercurrent speeds $V(r)$. Although based upon semiclassical theory, during development of our technique within that approximation[24] we demonstrated quantitatively correct supercurrent speeds $\boldsymbol{V}$ at $R \gtrsim 2\xi$, where $\xi$ is the superconductor's coherence length and $R$ is the scale associated with supercurrent variations[24]. For $Bi_2Sr_2CaCu_2O_8$, direct visualization reveals a true vortex core (not a vortex halo) where the order parameter reaches zero with a coherence length $\xi \approx 0.7\,nm$, while the typical region surrounded by a supercurrent (Figs. 2, 3, 4 below) has diameter $d \approx 3\,nm$ so that the semiclassical approximation appears not invalid. Thus, supercurrent speeds $V(\boldsymbol{r}) \equiv |V(\boldsymbol{r})|$ up to 4000 m/s may, in principle, be visualized via distinct characteristics of the measured differential conductance $g(\boldsymbol{r},E)$ spectrum translated to $V(\boldsymbol{r})$ using the "Rosetta Stone" in Fig. 1. Here we exploit this new strategy to detect and visualize $\mathcal{T}$ $-$symmetry breaking supercurrents in the canonical $CuO_2$-based superconductor, $Bi_2Sr_2CaCu_2O_8$.

High-data-volume spectroscopic imaging scanning tunneling microscopy (SISTM) is used for visualization of the density-of-electronic-states $N(\boldsymbol{r},E)$. The $Bi_2Sr_2CaCu_2O_8$ samples $0.17 \lesssim p \lesssim 0.22$ are inserted into a cryogenic SISTM, cleaved at $T$<10 K, and inserted in the STM scan head (Methods A). The differential tunneling conductance spectrum $dI/dV(\boldsymbol{r},E) \equiv g(\boldsymbol{r},E) \propto N(\boldsymbol{r},E)$ allows atomic-scale imaging of $N(\boldsymbol{r},E)$, whose severe chaoticity is well known[32,33,34], rendering subtle signatures of circulating supercurrents (e.g. Figs. 1f-i) impossible to discern. Indeed, analysis of such spectra at all spatial locations $\boldsymbol{r}$ using a $d_{x^2-y^2}$ version of Eqn. 1 and for all possible $V(\boldsymbol{r})$, has long proven numerically intractable (Methods B).

Today, however, artificial neural network (ANN) analysis of variegated high-volume experimental data sets[25] represents a promising opportunity to detect and visualize cuprate supercurrents. Consider the sequence of spectra in $d$ −wave superconductor in Figs. 1f-i: clearly, to detect such subtle phenomena one needs to evaluate all $g(\boldsymbol{r}, E)$ in terms of the energy gap $\Delta(\boldsymbol{r})$, the supercurrent speed $V(\boldsymbol{r})$ and the supercurrent trajectory $L(\boldsymbol{r})$. We have developed a dual-ANN system dubbed VMAP for that specific task (Methods C1). VMAP consists of two modules: MAIN and COGNATE, with MAIN being the initial analysis step and COGNATE being the final precision step. To train VMAP we use a $Bi_2Sr_2CaCu_2O_{8+x}$ normal state band dispersion:

$$E(k) = \mu - 2t_1(cosk_x + cosk_y) + 4t_2 cosk_x \cos k_y - 2t_3(cos2k_x + cos2k_y) \qquad (2)$$

where $\mu$ is the chemical potential, $t_1$, $t_2$, and $t_3$ are electron hopping parameters. The superconducting gap function is:

$$\Delta(k) = \frac{\Delta}{2}(cosk_x - cosk_y) \qquad (3)$$

where $\Delta$ is the maximum gap amplitude. For a supercurrent with velocity $\boldsymbol{V} = (V_x, V_y)$, the Bogoliubov band dispersion is modified by the Galilean boost $E_G(k) = \hbar(V_x k_x + V_y k_y)$ yielding a Bogoliubov-DeGennes Hamiltonian[35]:

$$H_{GB} = \begin{bmatrix} E(k) + E_G(k) & \Delta(\mathrm{k}) \\ \Delta(k)^* & -E(-k) + E_G(k) \end{bmatrix} \qquad (4)$$

The Green's function for eigenstates of $H_{GB}$ is:

$$\mathrm{G}(\omega, k, \Delta, V_x, V_y) = [\omega I - H_{GB}]^{-1} = A \begin{bmatrix} \omega + E(-k) - E_G(k) & \mathit{\Delta}(k) \\ \mathit{\Delta}(k)^* & \omega - E(k) - E_G(k) \end{bmatrix} \qquad (5)$$

where

$$A = \frac{1}{(\omega - E(k) - E_G(k))(\omega + E(-k) - E_G(k)) - \Delta^2(k)} \qquad (6)$$

We employ an energy $\hbar\omega = E + i\Gamma$ with a quasiparticle lifetime parameter $\hbar/\Gamma$[34]. The density of states $N(E)$ can then be derived from the imaginary part of the Green's function as:

$$N(E, \Delta, V_x, V_y, \Gamma) \equiv -\frac{1}{2\pi} Im\left[\sum_k Tr(G(\omega, k, \Delta, V_x, V_y))\right] \qquad (7)$$

Quantitively accurate simulations of the Galilean boost effects in Eqn. 7 for the $d_{x^2-y^2}$ superconductivity of $Bi_2Sr_2CaCu_2O_8$ are shown in Figs. 1f-i for various supercurrent velocities $\boldsymbol{V}(\boldsymbol{r})$ at angles 0, $\pi/8$, $\pi/4$, and $3\pi/8$ relative to the order parameter. But, from symmetry, Eqn. 7 results in 8 supercurrent velocity vectors $(\pm V_x, \pm V_y), (\pm V_y, \pm V_x)$ all having the same $N(E)$ (Methods C). Thus we introduce the quantities $V_H = Max(|V_x|, |V_y|)$ and $V_L = Min(|V_x|, |V_y|)$ being the maximal and minimal components of supercurrent velocity vector $\boldsymbol{V}$. For training the MAIN and COGNATE models, we generate 3.5 million $N(E, \Delta, V_H, V_L, \Gamma)$ spectra using Eqn. 7 with energy in the range $[-100\ \text{meV},\ 100\ \text{meV}]$, $\Delta$ in the range $[0\ \text{meV},\ 80\ \text{meV}]$, $\Gamma$ in the range $[0\ \text{meV},\ 10\ \text{meV}]$, $V_H$ and $V_L$ in the range $[0,\ 5000\ \text{m/s}]$ (Methods C). Tight-binding parameters for $E(k)$ are from angle resolved photoemission studies[36]. This synthetic dataset of 3.5 million simulated $N(E, \Delta, V_H, V_L, \Gamma)$ is used to train both the MAIN and COGNATE models of VMAP.

Experimentally, $g(E)$ spectra are typically of the form $g(E) = (a + bE) \times N(E, \Delta, V_H, V_L, \Gamma)$, where both $a$ and $b$ are small parameters and unavoidable electronic noise also exists. Thus, Gaussian noise and a linear-$E$ background were added as augmentation to each Eqn. 7 simulated $N(E)$ when training MAIN, with the noise level consistent with that measured in experimental $g(\boldsymbol{r}, E)$ data (Methods C3). In operation, the MAIN module determines four primary parameters $\Delta(\boldsymbol{r})$, $V_H(\boldsymbol{r})$, $V_L(\boldsymbol{r})$, and $\Gamma(\boldsymbol{r})$, with the consequent supercurrent speed being $V(\boldsymbol{r}) \equiv \sqrt{V_H^2(\boldsymbol{r}) + V_L^2(\boldsymbol{r})}$ (Methods C4). To establish best correspondence of VMAP determined $N(\boldsymbol{r}, E)$ with each experimental spectrum $g(\boldsymbol{r}, E)$, the parameters $\Delta(\boldsymbol{r})$, $V_H(\boldsymbol{r})$, $V_L(\boldsymbol{r})$, and $\Gamma(\boldsymbol{r})$ are used by the COGNATE module. This back-propagates gradients of the physical parameters efficiently, allowing precise evaluation of $N(\boldsymbol{r}, E)$ (Methods C5). Ultimately, the quality of fit parameter $R^2(\boldsymbol{r})$ is evaluated by comparing each measured $g(\boldsymbol{r}, E)$ to its VMAP reported $N(\boldsymbol{r}, E)$ spectrum. Our MAIN+COGNATE process increases the FOV-average to $R^2 \gtrsim 0.95$ in general, as demonstrated directly using data videos in Supplementary. Each empirical data set consists of a $Bi_2Sr_2CaCu_2O_8$ topographic image $T(\boldsymbol{r})$ with a simultaneous and atomically registered $g(\boldsymbol{r},E)$ (Methods G). VMAP analysis of $g(\boldsymbol{r},E)$ then yields images of the superconducting energy gap $\Delta(\boldsymbol{r})$ and of supercurrents $V(\boldsymbol{r}) \equiv \sqrt{V_H^2(\boldsymbol{r}) + V_L^2(\boldsymbol{r})}$. The validity of determination of $N(\boldsymbol{r},E)$, and thus of $\Delta(\boldsymbol{r})$ and $V(\boldsymbol{r})$, is exemplified directly in the Supplementary using three-panel videos of simultaneous $g(\boldsymbol{r},E)$: $N(\boldsymbol{r},E)$: $|g(\boldsymbol{r},E) - N(\boldsymbol{r},E)|$ which demonstrate high-quality VMAP fits to $g(\boldsymbol{r},E)$, throughout. Using a second ANN named V0MAP (Methods D) for which all training spectra are set at $\boldsymbol{V} = 0$, we demonstrate by comparing $R^2(\boldsymbol{r})$ for all $N(\boldsymbol{r},E)$: $g(\boldsymbol{r},E)$ correspondences from both ANNs, that supercurrent speeds $V \lesssim 1250$ m/s are indistinguishable from zero when using VMAP (Methods E). Nevertheless, the $V \gtrsim 1250$ m/s supercurrent trajectories in $\boldsymbol{V}(\boldsymbol{r})$ can be visualized directly using

$$L(x,y) = -\frac{\left[1+\alpha\left(\frac{\partial V}{\partial x}\right)^2\right]\alpha\frac{\partial^2 V}{\partial y^2} - 2\alpha^2\left(\frac{\partial V}{\partial x}\right)\left(\frac{\partial V}{\partial y}\right)\frac{\partial^2 V}{\partial x \partial y} + \left[1+\alpha\left(\frac{\partial V}{\partial y}\right)^2\right]\alpha\frac{\partial^2 V}{\partial x^2}}{\left[1+\alpha\left(\frac{\partial V}{\partial x}\right)^2 + \alpha\left(\frac{\partial V}{\partial y}\right)^2\right]^{\frac{3}{2}}} \qquad (8)$$

(Methods F) designed to identify the spatial locations where intense supercurrents appear. Finally, for $p < 0.17$, the VMAP system does not return valid results because the ANN training model (Eqns. 2-7) is demonstrably untenable for underdoped cuprates.

Exploiting these ANN techniques, we studied the $p$-dependence of $\Delta(\boldsymbol{r})$, $V_H(\boldsymbol{r})$, $V_L(\boldsymbol{r})$, and $\Gamma(\boldsymbol{r})$ phenomena. Focusing on $0.17 \lesssim p \lesssim 0.22$ with relevant $T(\boldsymbol{r})$ shown Methods G, the results of imaging $\Delta(\boldsymbol{r})$ by VMAP are shown in Figs. 2a-c. These are highly consistent with conventional heterogeneous $\Delta(\boldsymbol{r})$ studies[32,33,34]. VMAP imaging of $V(\boldsymbol{r})$ in the FOV associated with each $\Delta(\boldsymbol{r})$ at left, generates the results shown in Figs. 2d-f. These images are remarkable, revealing the direct signature of persistently flowing supercurrents at nanometer scale in a canonical $d$ −wave cuprate superconductor, with speeds reaching 4000 m/s. The statistical repeatability of $V(\boldsymbol{r})$ phenomena is manifest. Separately visualized $V^+(\boldsymbol{r})$ using $g(r,E>0)$ and $V^-(\boldsymbol{r})$ using $g(r,E<0)$ reveals them indistinguishable, consistent with the Hamiltonian in Eqn. 4 (Methods H) and demonstrating the insensitivity to noise. A network of intense supercurrents with patches of mean diameter 2.7±0.9 nm (Methods I) is revealed in all cases.

The spatial resolution is fixed at a typical 0.2 nm pixel size, which is far below the observed patch size. The supercurrents are most easily visualized in Figs. 2g-i, showing the network of supercurrent loop trajectories $L(\boldsymbol{r})$ associated with each panel at left.

For Galilean boosts to $N(E)$ (Fig. 1d) very specific characteristics should occur for different supercurrent directions relative to the $d_{x^2-y^2}$ symmetry order parameter (Figs. 1f-i and Methods J). In Fig. 3a we explore these characteristics for numerous supercurrent loop trajectories identified by $L(\boldsymbol{r})$ imaging. The color of dots indicates different directions relative to the order parameter. Figure 3b shows the $R^2(\boldsymbol{r})$ evaluated by VMAP in the same FOV of Fig. 3a. In Fig. 3c we show explicitly the predicted sequence of spectra for a circular supercurrent loop of either chirality. Figures 3d-i then report the sequence of measured $g(\mathrm{r}, E)$ spectra on typical loop trajectories labeled $1^{\#}$-$6^{\#}$ in Fig. 3a. Their detailed correspondence with Fig. 3c underlines the validation of supercurrent speed $V(\boldsymbol{r})$ imaging (Figs. 3d-i) using VMAP. Methods K and Extended Data Fig. 10 show the equivalent correspondence as a function of $p$.

Supercurrent loops ought to exhibit yet another key characteristic, that of satisfying a continuity equation. We thus consider junctions where three supercurrents meet, and examine if quantitively measured supercurrents entering/leaving the junction satisfy Kirchhoff's law. We mask the $V(r)$ images (e.g. Figs. 2d-f) by simultaneous images of loop trajectories $L(r)$ (e.g. Figs. 2g-i respectively), to produce a new image $LV(r)$ focusing on the supercurrent speeds along each trajectory in the network. A typical example is shown in Fig. 4a, revealing many tri-junctions and the measured supercurrent speeds approaching and departing them. Amazingly, the sum of two supercurrents adds to the value of the third in all well-defined tri-junctions detected within $LV(r)$ for all samples. Typical examples of this significant phenomenon are shown in Figs. 4b-h. More generally, a global supercurrent continuity constraint is explored using $S_i \equiv \sum_j V_j$, where $j$ are the four $x/y$ nearest neighbors of site $i$ and $V_j$ values can be any of $(\pm V_H, \pm V_L)$. Evaluating continuity, the objective is to determine if the sum of supercurrents entering and leaving any given pixel can add to 0, requiring a search for the minimum value of $S_i$ labeled $S(r)$. (Methods L). The measured $S(r)$ results are shown in Figs. 4i-k and comparable quantitively with the supercurrent speeds from which they are derived in Figs. 2d-f respectively. Here the mean value of $S(r)$ is below 20 m/s although supercurrent speeds in Figs. 2d-f reach up to 4000 m/s. This implies that supercurrent continuity is maintained throughout the tortuously heterogeneous $V(\boldsymbol{r})$ images (Figs. 2d-f) to within approximately 1%. Throughout Fig. 4 one emphasizes that, if the VMAP ANN were dominated by poor SNR, or incorrect evaluation of $V(\boldsymbol{r})$ from $N(r, E)$, or conflation of heterogeneities from another degree of freedom (e.g. $\Delta(r)$ or $\Gamma(r)$) into $V(r)$ , there would be no reason whatsoever for $V(\boldsymbol{r})$ to satisfy either a global continuity equation in $S(r)$ (Figs. 4i-k) or the local Kirchhoff's law constraint (Figs. 4b-h). Hence, these results sustain VMAP derived $V(\boldsymbol{r})$ as images of persistent supercurrents.

The long-standing challenge has been to identify and understand $\mathcal{T}-$symmetry breaking in cuprates[2-23]. With respect to disordered intra-unit-cell loop-current systems sustaining domain wall supercurrents[17,18], our results are perhaps not relevant because we focus on overdoped cuprates. For such samples, a nanoscale network of supercurrents induced by disorder in $d-$wave superconductors[12,19-22] seems more consistent with direct supercurrent visualization (Figs. 2-4). Revealed is a novel superconductive state heterogeneously breaking $\mathcal{T}$ −symmetry somewhat akin to a spin-glass phase, but consisting of a disordered network of supercurrent loops, whose chirality cannot yet be determined. These networks exhibit speeds of several thousand m/s. While subsurface magnetic impurity effects can be ruled out on the basis of their minimal induced supercurrents (Methods I), spontaneous loop currents on these scales were predicted by both Refs. 21 and 22, using mean-field simulations for a disordered $d-$wave superconductive state. Within closed loops of approximate diameter 3 nm, the observed supercurrents satisfy the circulation quantization constraint. Reference 22 predicts correct supercurrent speeds, loop sizes, and formation of supercurrent 'mesas', consistent with observations. However, different microscopic origins are proposed for the disordered loop current state in Ref. 21 and Ref. 22. At present, while it is possible to sustain the physical validity of this general class of supercurrent $\mathcal{T}-$symmetry breaking theories, it is not possible to sharply distinguish between them. Theoretical advances will be required to achieve that aim.

Nevertheless, within the context of Eqns. 4-7 the empirics in Figs. 2-4 provide a radically new perspective on the physics of $\mathcal{T}-$symmetry breaking in highly doped superconducting cuprates. First, an hypothetical term $E_G(k)$ in Eqn. 4 that is consistent with the experiments (Figs. 2-4) has few if any other possible causes except a Galilean energy boost due to supercurrents. Second, when we separately visualize $V^{+}(\boldsymbol{r})$ using the positive bias $g(r, E > 0)$ data and $V^{-}(\boldsymbol{r})$ using the negative bias $g(r, E < 0)$ data, they are practically indistinguishable (Methods H) emphasizing both the particle-hole symmetry of $V(\boldsymbol{r})$ and the adequacy of the experimental signal to noise ratio. Third, the correspondence between $g(\boldsymbol{r}, E)$ spectra measured along the trajectory of supercurrent loops (Fig. 3) to the supercurrent "Rosetta Stone" (Figs. 1e-i) provides substantiation of VMAP's supercurrent speed $V(\boldsymbol{r})$ visualization. Fourth, circulation is observably quantized around close supercurrent loops. Fifth, from theoretic models with Fermi velocity $V_F \approx 3.5 \times 10^5$ m/s, a typical disordered $d-$wave supercurrent velocity is predicted[22] to be $V \approx 0.01V_F$ , in good agreement observations (Fig. 2). Sixth, approximate estimates of the absolute value of loop supercurrents $I$ based on the measured $V(\boldsymbol{r})$ and estimated electron-pair density, yield $I{\sim}10^{-7}$ A, so that that for ~1.5 nm radius domains the magnetic field at center is $|B|{\sim}0.1\ mT$ (Methods I). This is in reasonable agreement with reported spontaneous fields of $\mathcal{T}-$symmetry breaking in cuprates. Finally, the supercurrent images $V(\boldsymbol{r})$ are consistent with global continuity constraints (Figs. 4i-k) and locally satisfy Kirchhoff's law at all valid tri-junctions (Figs. 4b-h). Thus, Figs. 2-4 provide novel empirics on $\mathcal{T}-$symmetry breaking in cuprates. They also emphasized the accelerating utility ANN techniques for physics discovery[25] in high data volume quantum matter research. Ultimately, this general capability to visualize supercurrent loops may be employed to resolve outstanding mysteries of $\mathcal{T}-$symmetry breaking in a variety of other superconductor classes[12].

**Acknowledgements**: We acknowledge and thank C. Varma for key motivation and guidance for this project. We thank N. Kennedy and A. Samoilenka for key advice.

**Funding:** P.J.H acknowledges support from the Leverhulme Foundation and NSF grant NSF-DMR-2231821. Shuqiu W. acknowledges support from the Engineering and Physical Sciences Research Council (EPSRC) under Award EP/Z53660X/1 and the support from the Royal Academy of Engineering / Leverhulme Trust Research Fellowship. B.H. acknowledges support from Research Ireland under GOI Postdoctoral Fellowship GOIPD/2026/2219. K.Z. acknowledges EU Horizon Europe MSCA funding under grant agreement No. 101203931 (QUASAR). J.C.S.D. acknowledges support from the Moore Foundation's EPiQS Initiative through Grant GBMF9457. B.H. and J.C.S.D. acknowledge support from the UK Royal Society under Award R64897. Siyuan. W., H.Z., C.D., B.H, J.W., and J.C.S.D. acknowledge support from Research Ireland under Awards SFI 17/RP/5445 and RI 25/RP-2TF/13559

**Competing interest:** The authors declare no competing interests.

**Contact details for corresponding authors:** korbin.hu@gmail.com (Bin Hu), jcseamusdavis@gmail.com (J.C. Séamus Davis), pjh@phys.ufl.edu (Peter J. Hirschfeld).

## Main Figure Legends

**Figure 1 $\boldsymbol{d}-$ wave Galilean boost to Bogoliubov density of states from supercurrents**

**a.** Schematic diagram of domain wall supercurrents within disordered intra-unit-cell loop-current systems.

**b.** Schematic diagram of nanoscale superconductive phase-heterogeneity induced supercurrents for $d_{x^2-y^2}$ symmetry superconductors.

**c.** Schematic diagram of strong impurity scattering induced $d + is$ supercurrents, for $d_{x^2-y^2}$ symmetry superconductors.

**d.** Schematic of the Galilean boost effect on Bogoliubov quasiparticle energy and consequent alteration of $N(E)$, due to supercurrents with speed $\vec{V} > 0$.

**e.** Schematic of a supercurrent loop oriented relative to the order parameter of a $d_{x^2-y^2}$superconductor. The predicted $N(E)$ is different for each supercurrent direction and speed as shown in panels f-i. This is the "Rosetta Stone" translating each $N(E)$ spectrum to a specific speed V, and froms the training set for VMAP.

**f.** $N(E)$ spectra from Eqns. 2-7 for supercurrents with different speed $V$ at an angle $0$ to $d - \mathrm{wave}$ order parameter.

**g.** $N(E)$ spectra from Eqns. 2-7 for supercurrents with different speed $V$ at an angle $\pi/8$ to $d - \mathrm{wave}$ order parameter.

**h.** $N(E)$ spectra from Eqns. 2-7 for supercurrents with different speed $V$ at an angle $\pi/4$ to $d - \mathrm{wave}$ order parameter.

**i.** $N(E)$ spectra from Eqns. 2-7 for supercurrents with different speed $V$ at an angle 3$\pi/8$ to $d$ –wave order parameter.

**Figure 2 VMAP determination of $\Delta(\boldsymbol{r})$, $\boldsymbol{V}(\boldsymbol{r})$ and $\boldsymbol{L}(\boldsymbol{r})$ versus hole density $\boldsymbol{p}$**

**a-c** Imaging energy gaps $\Delta(\boldsymbol{r})$ versus hole-density $p$ using VMAP. Relevant topographies are shown in Methods G and Extended Data Fig. 6. Data demonstrating the overall quality of to $g(\boldsymbol{r}, \boldsymbol{E})$ by VMAP is presented using data videos in Supplementary.

**d-f** Imaging supercurrent speeds $V(\boldsymbol{r})$ in the same FOV as associated $\Delta(\boldsymbol{r})$ panel to left, versus hole-density $p$ using VMAP.

**g-i** Imaging the network of supercurrent loops $L(\boldsymbol{r})$ in the same FOV as associated $\Delta(\boldsymbol{r})$ panel at left, versus hole-density $p$ using VMAP, where $\sigma_r$=4 Å, $\alpha$=0.30 (ns)$^2$.

**Figure 3 Individual supercurrent loops as identified by VMAP**

**a.** Supercurrent speeds on designated current loops 1#-6# in $L(\boldsymbol{r})$. Color code is as in Fig. 1e, designating distinct directions of supercurrent. There is an ambiguity as to whether the supercurrents are flowing clockwise or anti clockwise.

**b.** Goodness of fit $R^2(r)$ of all VMAP $N(r, E)$ to measured $g(r, E)$, simultaneously measured with **a**. The normalized cross correlation coefficient between **a** and **b** is 0.03 demonstrating the typical result that there is virtually no spatial correspondence between $R^2(r)$ and $L(r)$ so that the rare examples of deficient fitting do not falsely create the images of supercurrent speeds.

**c.** Sequence of theoretical $N(r, E)$ spectra on the trajectory of a supercurrent loop. Color code is as in Fig. 1e, designating distinct directions of supercurrent.

**d.** Sequence of $g(r, E)$ spectra on the trajectory of loop 1#.

**e.** Sequence of $g(r, E)$ spectra on the trajectory of loop 2#.

**f.** Sequence of $g(r, E)$ spectra on the trajectory of loop 3#.

**g.** Sequence of $g(r, E)$ spectra on the trajectory of loop 4#.

**h.** Sequence of $g(r, E)$ spectra on the trajectory of loop 5#.

**i.** Sequence of $g(r, E)$ spectra on the trajectory of loop 6#. The spectra in **d-i** are vertically shifted for clarity and have the corresponding speed listed in m/s.

**Figure 4 Kirchhoff constraint exhibited by supercurrent loops**

**a.** To identify both the supercurrent speeds and trajectories we masked VMAP measured $V(r)$ by the $L(r) > 0.5$ s/m trajectory image with typical $LV(r)$ results shown here. The $L(r) > 0.5$ s/m cut-off is strict and current can flow.

**b.** A typical tri-junction showing the VMAP quantified supercurrent speeds crossing the dashed line on approaching/departing all three channels. Two of these values add to the third as expected for Kirchhoff's law.

**c**-**g** Typical examples of the key fact that this Kirchhoff's law criterion is satisfied at all well-defined tri-junctions throughout all samples studied.

**i**-**k** The $S(r)$ images obtained from $V(r)$ in Fig. 2. A minimal boxcar averaging was applied to $V(r)$ before calculating the $LV(r)$ and $S(r)$.

# Fig. 1

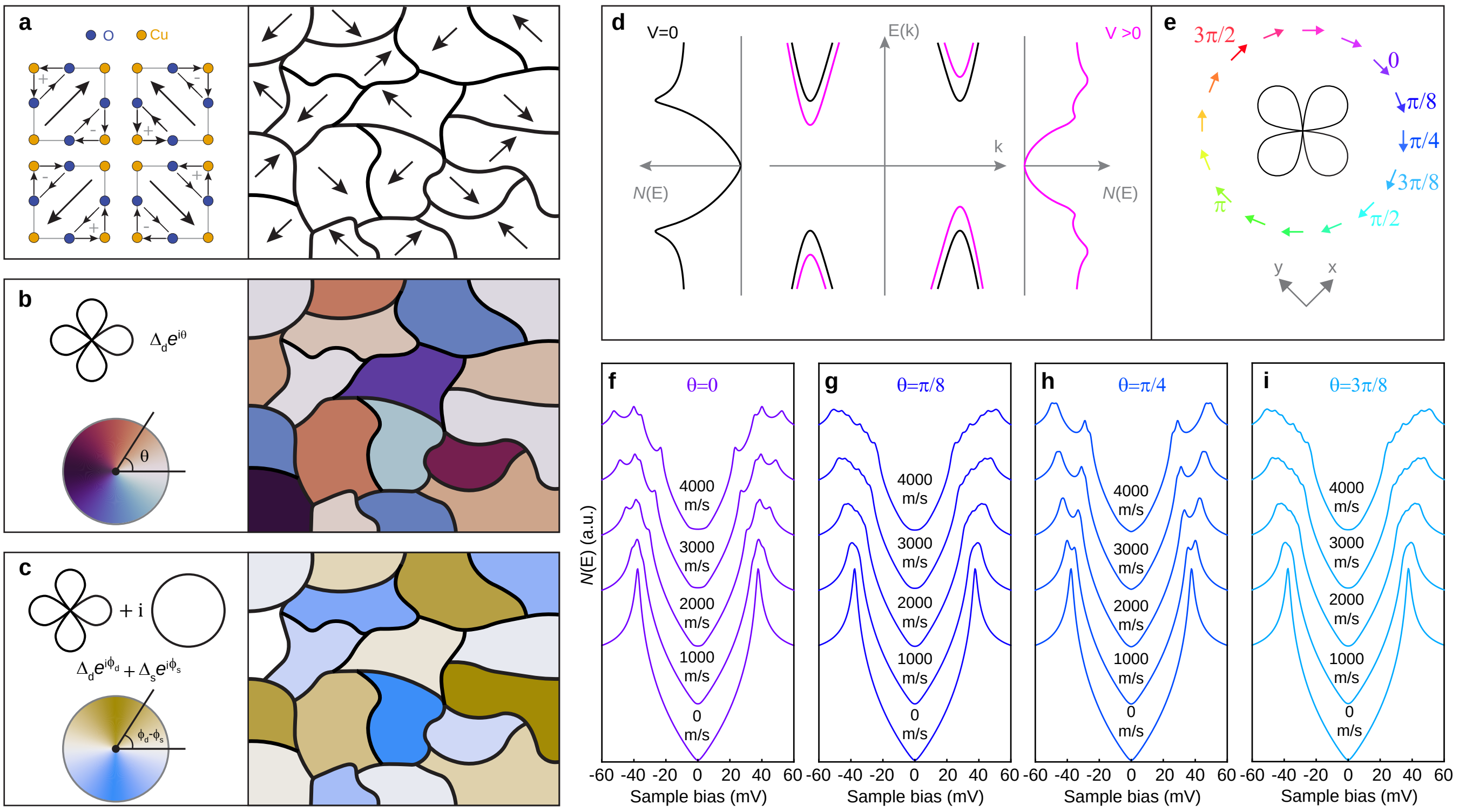

a
O
Cu
b
$\Delta_d e^{i\theta}$
θ
c
+ i
$\Delta_d e^{i\phi_d} + \Delta_s e^{i\phi_s}$
$\phi_d$-$\phi_s$
d
V=0
E(k)
V >0
k
N(E)
N(E)
e
3π/2
0
π/8
π/4
3π/8
π/2
π
y
x
f
θ=0
g
θ=π/8
h
θ=π/4
i
θ=3π/8
4000 m/s
3000 m/s
2000 m/s
1000 m/s
0 m/s
N(E) (a.u.)
-60 -40 -20 0 20 40 60
Sample bias (mV)

Fig. 2

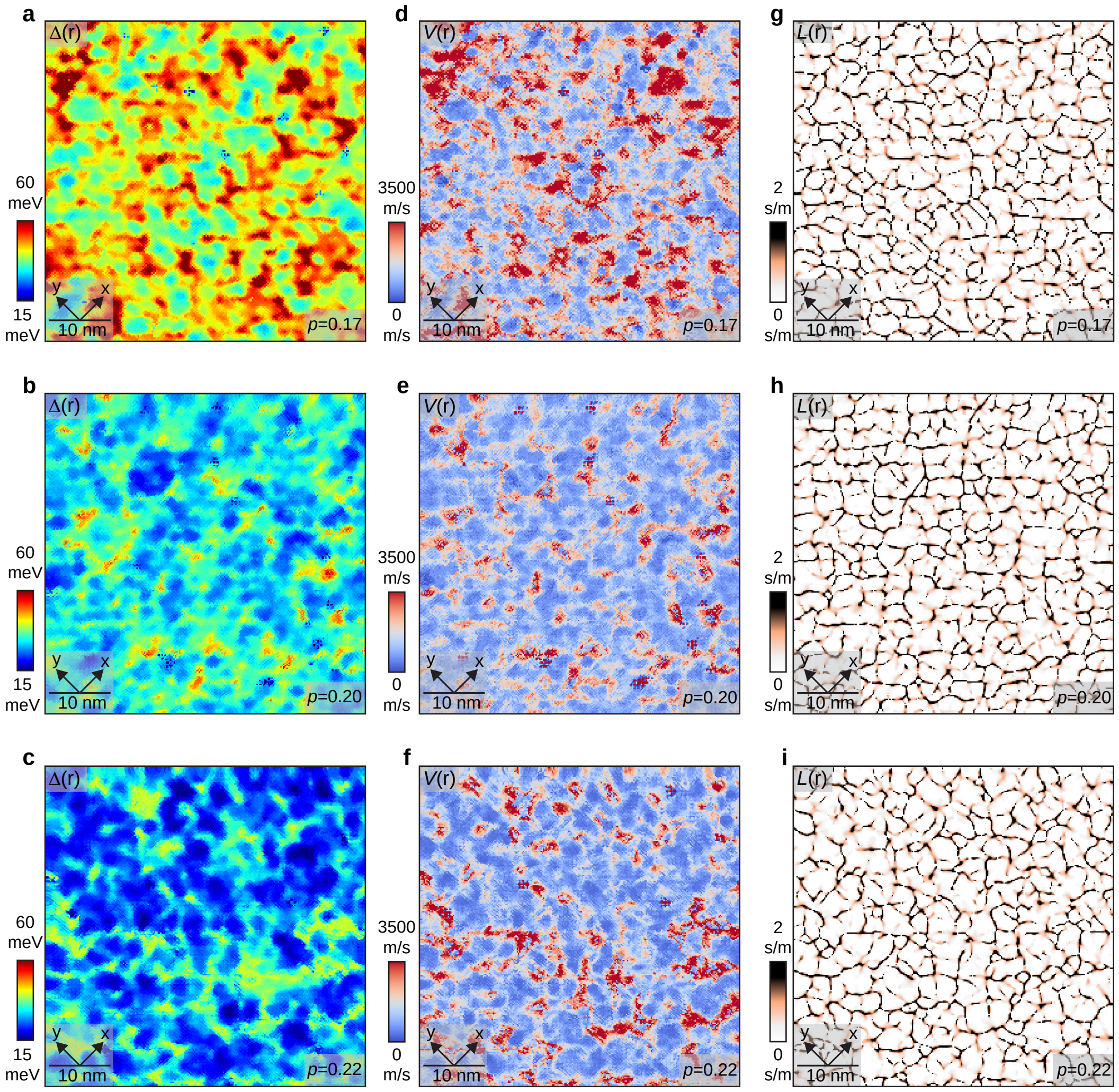

a
Δ(r)
60 meV
15 meV
10 nm
p=0.17
b
Δ(r)
p=0.20
c
Δ(r)
p=0.22
d
V(r)
3500 m/s
0 m/s
p=0.17
e
V(r)
p=0.20
f
V(r)
p=0.22
g
L(r)
2 s/m
0 s/m
p=0.17
h
L(r)
p=0.20
i
L(r)
p=0.22
y
x

# Fig. 3

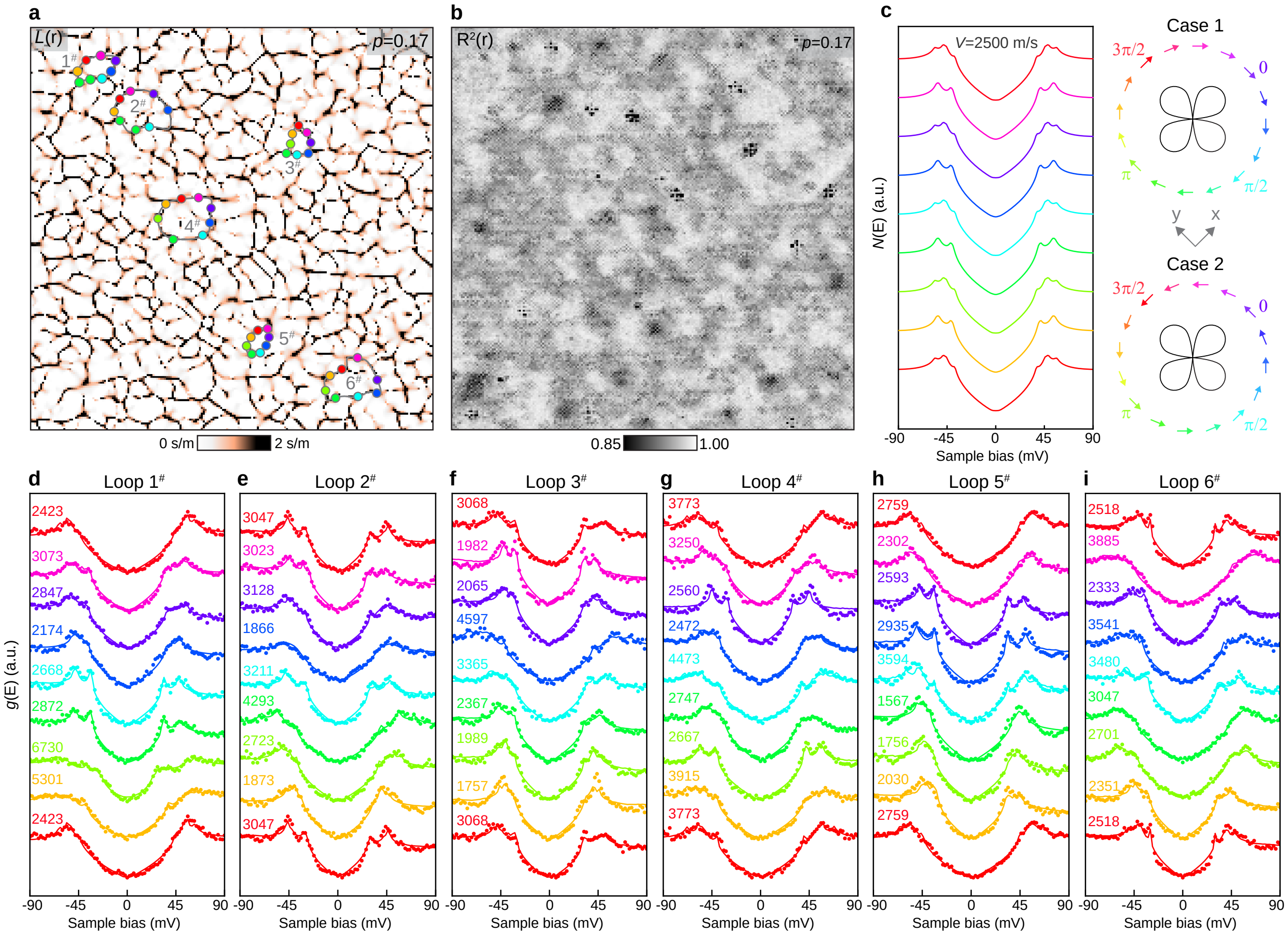
a
L(r)
p=0.17
0 s/m
2 s/m
b
R²(r)
p=0.17
0.85
1.00
c
V=2500 m/s
N(E) (a.u.)
Sample bias (mV)
Case 1
Case 2
3π/2
0
π/2
π
y
x
d
Loop 1#
e
Loop 2#
f
Loop 3#
g
Loop 4#
h
Loop 5#
i
Loop 6#
g(E) (a.u.)
Sample bias (mV)
2423
3073
2847
2174
2668
2872
6730
5301
2423
3047
3023
3128
1866
3211
4293
2723
1873
3047
3068
1982
2065
4597
3365
2367
1989
1757
3068
3773
3250
2560
2472
4473
2747
2667
3915
3773
2759
2302
2593
2935
3594
1567
1756
2030
2759
2518
3885
2333
3541
3480
3047
2701
2351
2518

# Fig. 4

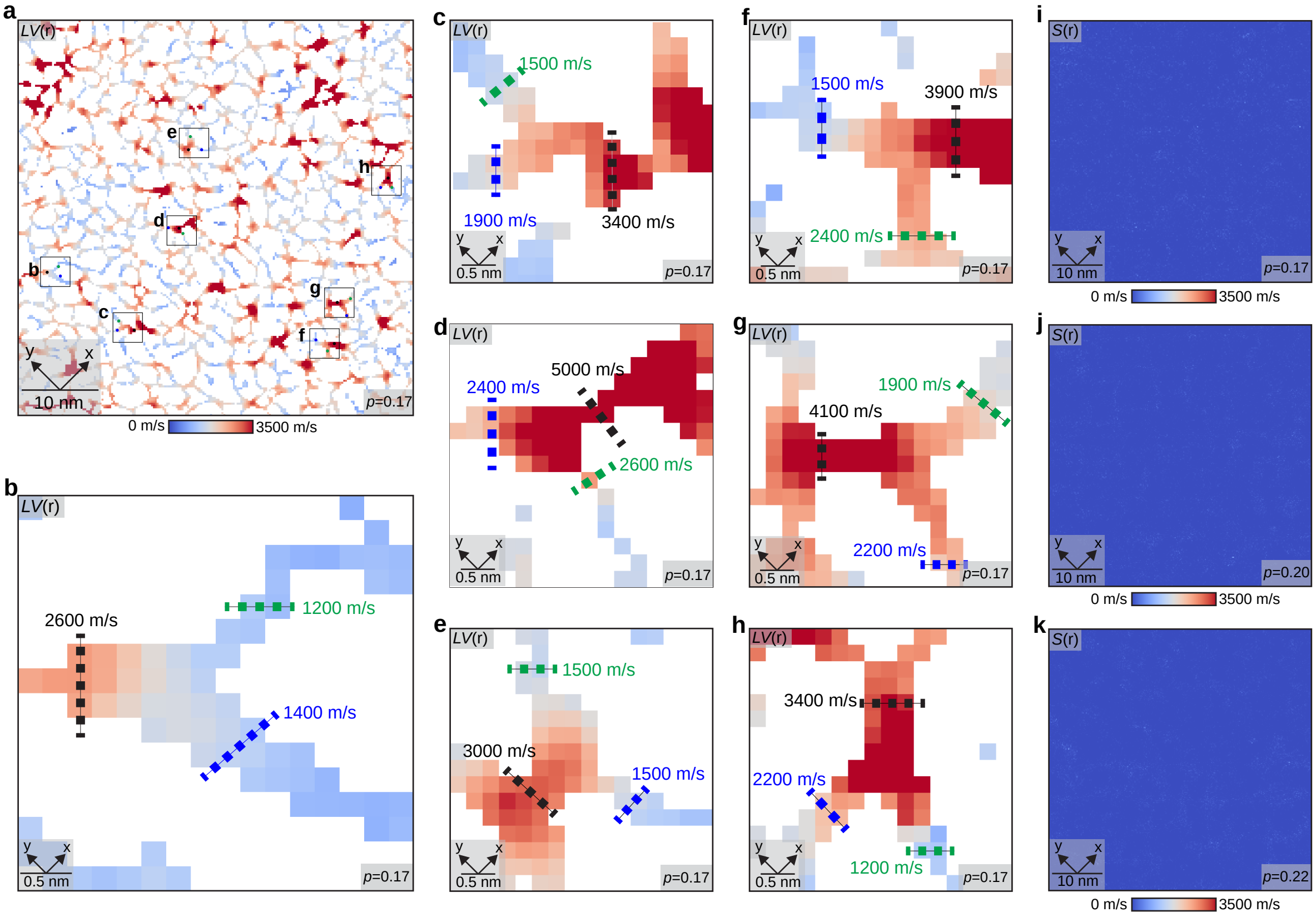

a
LV(r)
e
h
d
b
g
c
f
10 nm
p=0.17
0 m/s
3500 m/s
b
LV(r)
2600 m/s
1200 m/s
1400 m/s
0.5 nm
p=0.17
c
LV(r)
1500 m/s
1900 m/s
3400 m/s
0.5 nm
p=0.17
d
LV(r)
5000 m/s
2400 m/s
2600 m/s
0.5 nm
p=0.17
e
LV(r)
1500 m/s
3000 m/s
1500 m/s
0.5 nm
p=0.17
f
LV(r)
1500 m/s
3900 m/s
2400 m/s
0.5 nm
p=0.17
g
LV(r)
1900 m/s
4100 m/s
2200 m/s
0.5 nm
p=0.17
h
LV(r)
3400 m/s
2200 m/s
1200 m/s
0.5 nm
p=0.17
i
S(r)
10 nm
p=0.17
0 m/s
3500 m/s
j
S(r)
10 nm
p=0.20
0 m/s
3500 m/s
k
S(r)
10 nm
p=0.22
0 m/s
3500 m/s

## Methods

### A. Materials and measurements

Single crystals of $Bi_2Sr_2CaCu_2O_8$ were grown by the traveling solvent-floating-zone (TSFZ) method. The spectroscopic imaging scanning tunneling microscope (SISTM) measurements of $Bi_2Sr_2CaCu_2O_8$ crystals were performed on a custom instrument at 4.2 K. All $Bi_2Sr_2CaCu_2O_8$ samples were cleaved in situ in cryogenic ultra-high vacuum at $T$<10 K and then immediately transferred into the STM head. All SISTM measurements were conducted using a W scan tip, well characterized on a Au single crystal, before conducting measurements on $Bi_2Sr_2CaCu_2O_8$ samples.

### B. Method on supercurrent velocity analysis

The Galilean-boost energy $E_G(k) = \hbar(V_x k_x + V_y k_y)$ is $k$-dependent, while the density of states $N$(E) integrates over the whole BZ in $k$-space, meaning that 8 supercurrents result in the same $N$(E) : $\left(\pm V_x, \pm V_y\right), (\pm V_y, \pm V_x)$ adding difficulty to the analysis. Earlier, we demonstrated a supercurrent visualization method[24] using a reference dataset, where each of the experimental spectra $g$(E) is compared individually to all of the reference dataset $N$(E), and the maximal coefficient of determination $R^2$ produces the best matched parameters. However, with four-parameter $d$-wave supercurrent fitting as in this project, such a reference dataset would contain ~1.6 billion spectra. Hence this approach is inadequate for high data volume and heterogeneous cuprate $g(r, E)$ data sets of a disordered $d$-wave superconductor.

### C1. VMAP ANN architecture

To tackle the eight possible $N$(E) for each supercurrent velocity $\boldsymbol{V}$, we introduce two quantities, namely the higher (lower) vectorial components $V_H$ ($V_L$) of $\boldsymbol{V}$ such that $\boldsymbol{V} = (V_H\hat{x}, V_L\hat{y})$ or $\boldsymbol{V} = (V_L\hat{x}, V_H\hat{y})$. Thus

$$V_H = Max(|V_x|, |V_y|) \quad \text{(M1)}$$

$$V_L = Min(|V_x|, |V_y|) \quad \text{(M2)}$$

where $Max$ and $Min$ are referred to as taking the maximal and minimal values, respectively. For a given parameter set $(\Gamma, \Delta, V_H, V_L)$, the $N(E)$ within the energy range $[E_1, E_2]$ can then be sharply defined. We use this as the reference dataset randomly sampling these four parameters. Next, we introduced a dual-artificial-neural-network-based method, namely VMAP, comprising a MAIN and a COGNATE. In practice, 3.5 million references $N(E)$ curves are sufficient to train VMAP.

For MAIN, we aim to take the measured $g(E)$ spectra as input and the four parameters $(\Gamma, \Delta, V_H, V_L)$ as output. However, this is a "one-to-many" mapping task, especially when the signal-to-noise ratio of input curve is inadequate, in which multiple parameter sets $(\Gamma, \Delta, V_H, V_L)$ can produce similar $N(E)$ curve. To improve the performance of MAIN, we design a COGNATE that identifies the $N(E)$ spectra from the MAIN output parameter set that corresponds best to the input $g(E)$.

The COGNATE adopts a transformer-like architecture[37] (Extended Data Fig. 1a). Each of the four parameters $(\Gamma, \Delta, V_H, V_L)$ is projected into the hidden space by an independent fully connected (FC) layer, and the four resulting representations are then fed into the self-attention block (Extended Data Fig. 1c). The energies corresponding to the sample bias in the experimental $g(E)$ spectra are projected into the hidden space via an FC layer as well with one energy per token and then used as input to self-attention. In the cross-attention section, the energy representations serve as queries, while the encoded parameter representations provide both keys and values. This decoder block (dashed rectangle in Extended Data Fig. 1a) is repeated four times. Each block comprises a self-attention block (SAB, $SAB(x) = x + SA(\text{norm}(x))$, where $SA$ is computing the scaled dot-product attention) over the projected energy, cross-attention, and a feedforward neural-network (FFN). A final FC layer maps the decoded energy representations to the predicted density of states $N(E)$.

The MAIN module is designed primarily to extract the four parameters $(\Gamma, \Delta, V_H, V_L)$ from a given $g(E)$ spectrum where $E$ is given (Extended Data Fig. 1b). In reality, a linear background signal may be present in the experimental $g(V)$ spectra which are its data source. Thus, we make the augmentation before the training dataset goes into MAIN, that is:

$$N_{\mathrm{A}}(E, \varGamma, \varDelta, V_H, V_L, a, b) = (a + bE) \times N(\mathrm{E}, \varGamma, \varDelta, V_H, V_L) \quad \text{(M3)}$$

For the feature recognition, we use a Convolution Neural Network (CNN). The CNN is consistent of nine CNN blocks, where each of the CNN blocks includes two branches (Extended Data Fig. 1d). In the CNN block, the first branch starts with the $n \times 1$ learnable convolution layer connected with two $1 \times 1$ learnable convolution layers and the second branch is one $1 \times 1$ learnable convolution layer. In this work, the non-one kernel sizes are $n = 3, 5, 7, 9, 15, 21, 15, 7, 5$ for the nine CNN blocks in sequence. Yet CNN alone is insufficient for accurate four-parameter extraction. Therefore, we take a self-attention branch in parallel with the CNN branch, and the two branches are then concatenated. Ultimately, four FC heads output the four parameters: one each for $\varGamma$, $\varDelta$, $V_H$ and $V_L$ separately. The number of model parameters in either the COGNATE or the MAIN is less than one million.

**C2. Synthetic training-dataset generation of VMAP**

We generate 3.5 million $N(E)$ spectra using Eqn. 7, with energy in the range $[-100\ \mathrm{meV}, 100\ \mathrm{meV}]$, $\varDelta$ in the range $[0\ \mathrm{meV},\ 80\ \mathrm{meV}]$, $\varGamma$ in the range $[0\ \mathrm{meV},\ 10\ \mathrm{meV}]$, $V_H$ and $V_L$ in the range $[0\ \mathrm{m/s},\ 5000\ \mathrm{m/s}]$. The normal state tight-binding parameters are interpolated from Ref. 36 and presented in the Supplementary Table 1. To get a valid $N(E)$ curve, we use a dense $k$-mesh of size 13501×13501 and mask the effective $k$-points in the energy range $[-200\ \mathrm{meV},\ 200\ \mathrm{meV}]$. The same synthetic dataset is used to train both the COGNATE and the MAIN, where 3.15 million (90%) synthetic $N(E)$ curves were used for training and 0.35 million (10%) synthetic $N(E)$ curves were used for validation. We generate the synthetic dataset on 6×NVIDIA L40S GPUs, which takes approximately four days. Here, we also generate independent synthetic datasets for $Bi_2Sr_2CaCu_2O_8$ samples with different hole doping levels.

## C3. Synthetic dataset augmentation for training VMAP

To enhance the capability of the MAIN for interpreting experimental data which contains random noise, we implemented random noise augmentation on the synthetic dataset used by the MAIN for training. The noise $f_{noise}(E)$ is generated from a normal distribution generator (with mean value 0 and variance 1) to add random numbers at each energy point. The noise-augmented spectrum is $T_A(E) = N(E) + x f_{noise}(E) \times \overline{N(E)}$, where $N(E)$ is the synthetic spectrum with its mean value $\overline{N(E)}$, and $x$ is a scalar factor that represents the noise level. At each training epoch, an independent noise realization is used for each synthetic spectrum; the same spectrum experiences different noise augmentations across epochs. We limit $\ln(x)$ in a linear range from $\ln(0.005)$ to $\ln(0.15)$ with $x$ varying from 0.5% to 15% but centered at ~2.7%. Extended Data Fig. 2 shows the example synthetic spectra without noise (Extended Data Fig. 2a) and with noise from 1% to 4% (Extended Data Figs. 2b-d). With noise augmentation applied, the MAIN converges slower than trained on the clean synthetic dataset but achieves a better performance when evaluating the physical parameters of the experimental spectra. Noise augmentation is implemented only for the MAIN in this work.

## C4. VMAP Training

We use the Gaussian Error Linear Unit (GELU)[38] as the activation function and AdamW[39] optimizer throughout the VMAP. We also incorporate the position encoding[37] (PE) for the COGNATE and the self-attention branch of the MAIN. For the COGNATE, we use the mean squared error (MSE) loss function $L_{MSE}$ and take the coefficient of determination $R^2$ to evaluate the accuracy (Extended Data Fig. 2e).

$$L_{MSE} = \frac{1}{N}\sum_{i=1}^{N}(y_i - f_i)^2 \tag{M4}$$

where $y_i$ is the target spectrum and $f_i$ is the corresponding reconstructed curve at energy $i$. The COGNATE was trained for 100 epochs. For the MAIN, we use the mean absolute error (MAE) loss function $L_{MAE}$:

$$L_{MAE} = \frac{1}{N}\sum_{i=1}^{N} |p_i - l_i| \tag{M5}$$

where $p_i$, ($p_i = \Delta, \Gamma, V_H, V_L$), and $l_i$ are the predicted parameters and the corresponding true values, respectively. The accuracy of each parameter is defined as:

$$Accuracy(p_i) = 1 - \frac{\sum_s |p_{i,s} - l_{i,s}|}{\sum_s |l_{i,s}|} \tag{M6}$$

where the index $s$ runs over all samples in the validation dataset. The COGNATE converges quicker than the MAIN since the COGNATE aims to reconstruct the spectrum from the given physical parameters and was trained on the clean dataset. In practice, we train the COGNATE first, where the COGNATE shows excellent performance (Extended Data Fig. 2e). In addition, we introduce a Physics-Informed Neural Network (PINN) loss $L_{PINN}$ computed through the COGNATE to accelerate the convergence of the MAIN. The $L_{PINN}$ is the same as $L_{MSE}$ as in the COGNATE. The MAIN loss function is then the combined $L_{MSE}$ and $L_{PINN}$:

$$L_{MAIN} = (1 - w)L_{MAE} + wL_{PINN} \tag{M7}$$

where $w$ is a relative ratio and we take $w = 0.1$ when training the MAIN in this work. The MAIN is trained for 300 epochs, with the PINN loss active during epochs 20 to 150. When the PINN loss is active, the COGNATE only performs forward computation with frozen weights and bias. We do not use Eqn. 7 to compute the PINN loss directly, as the numerical $N$(E) calculations are computationally extensive. The validation loss and accuracy of the physical parameters of MAIN are shown in Extended Data Fig. 2f.

## C5. VMAP Parameter Determination

The experimental data $g(E)$ first is processed by MAIN, which outputs the four parameters ($\Gamma, \Delta, V_H, V_L$), but this is not the final output. Instead, we use the ($\Gamma, \Delta, V_H, V_L$) as input of the COGNATE, which generates the reconstructed spectrum $N_{COG}(E)$. By fitting the ratio $\frac{g(E)}{N_{COG}(E)}$ to a linear function $a + bE$, where $g(E)$ is the target experimental spectrum, $a$ and $b$ can be obtained. We then use the MSE loss in the refinement procedure:

$$L_{Fit} = \frac{1}{N}\sum_{i=1}^{N}(y_i(a + bE_i) - f_i)^2 \tag{M8}$$

In this scenario, the four physical parameters $(\Gamma, \Delta, V_H, V_L)$ receive gradients through the frozen COGNATE and are iteratively updated via the AdamW optimizer to minimize $L_{Fit}$. In this work, we perform 2000 optimization steps for each of the experimental spectrum and track the best parameters with the lowest $L_{Fit}$. The schematic of the refinement process is shown in Extended Data Fig. 3a and 3b.

The final six parameters are used to recompute the spectra based on Eqn. 7, and the fitting quality is defined by the coefficient of validity $R^2$:

$$R^2(\boldsymbol{r}) = 1 - \frac{\sum_{i=1}^{\mathrm{m}}[g(\boldsymbol{r}, E_i) - N(\boldsymbol{r}, E_i)]^2}{\sum_{i=1}^{m}[g(\boldsymbol{r}, E_i) - \bar{g}(\boldsymbol{r})]^2} \tag{M9}$$

in which $g(\boldsymbol{r}, E_i)$ is the measured differential conductance spectrum and $N(\boldsymbol{r}, E_i)$ is the numerical recomputed spectrum from Eqn. 7 based on the given six parameters, $\bar{g}(\boldsymbol{r})$ is the corresponding mean value of $g(\boldsymbol{r}, E_i)$. Extended Data Figs. 3c-f shows an example of the $R^2$ before and after the refinement procedure, from which the mean value of $R^2$ increased from 0.87 to 0.96.

**D. V0MAP**

In this work, the VMAP ANN has a lower limit on detectable supercurrent speeds. To quantify these limits statistically, we fix the $V_{\mathrm{H}} = V_{\mathrm{L}} = 0$ in a different ANN named V0MAP. V0MAP has identical ANN architecture as VMAP (Extended Data Fig. 1), but with the training supercurrents fixed at zero ($V_H = V_L = 0$). In terms of the Hamiltonian Eqn. 4 which generates the 3.5 million training spectra for V0MAP, the term $E_G \equiv \hbar\boldsymbol{k} \cdot \boldsymbol{V} = 0$. In this case, the V0MAP ANN searches for best fits using only $\Delta(r)$, $\Gamma(r)$ and the background $a(r) +$ $b(r)E$ as degrees of freedom. The $R^2$ comparison obtained from VMAP and V0MAP is presented in Extended Data Fig. 4.

**E. Indistinguishability from zero of small supercurrent velocities**

We compute the $R^2$ without supercurrent velocity and label it as $R^2(V=0)$, then the quantity $\delta R^2$ is defined as $\delta R^2 = R^2(V) - R^2(V=0)$, where $R^2(V)$ is computed with finite supercurrent velocity. Extended Data Figs. 4a-c show the supercurrent velocity-dependent $\delta R^2$, which is averaged at each velocity, indicating that the Galilean-boosted $N(E)$ has a better description on the $g(E)$ in general. Combined with the supercurrent velocity-dependent $\delta R^2$ at different doping levels, the conservative indistinguishability of small supercurrent speed is $\sim 1250\ (\pm 250)$ m/s in this work. It should be noted that the possible finer supercurrent structures for V<1250 m/s will be the focus of future research.

On the other hand, for clarity, we apply a cutoff $L(r) > 1$ mask onto the $\Delta(r)$ (Extended Data Figs. 5a-c), where the domains of low supercurrent are shown clearly. In Extended Data Figs. 5d-f, we plot the histograms of the masked superconductive gap $\Delta$ within those supercurrent-velocity indistinguishable regions, where the results are shown in Extended Data Figs. 5a-c, the average superconductive gap from those domains $\langle \Delta \rangle = 44(\pm 7)$ meV for hole doping $p$=0.17, and it decreases as the hole density increases, consistent with previous reportsand expectations.

**F. Visualizing Supercurrent Loop Network**

As an ancilla to VMAP for obtaining clearer images of the supercurrent loop network in $\boldsymbol{r}$-space, we adopt the two-dimensional curvature method on the supercurrent velocity maps $V(r)$[40]. For the square grid, where the pixel distances are equal along the $x$ and $y$ axes with the same units, the two-dimensional curvature image $L(x,y)$ of the target image $F = f(x,y)$ can be defined as (also see in Eqn. (8)):

$$L(x,y) = -\frac{\left[1+\alpha\left(\frac{\partial F}{\partial x}\right)^2\right]\alpha\frac{\partial^2 F}{\partial y^2} - 2\alpha^2\left(\frac{\partial F}{\partial x}\right)\left(\frac{\partial F}{\partial y}\right)\frac{\partial^2 F}{\partial x \partial y} + \left[1+\alpha\left(\frac{\partial F}{\partial y}\right)^2\right]\alpha\frac{\partial^2 F}{\partial x^2}}{\left[1+\alpha\left(\frac{\partial F}{\partial x}\right)^2+\alpha\left(\frac{\partial F}{\partial y}\right)^2\right]^{\frac{3}{2}}} \quad \text{(M10)}$$

where $\alpha$ is a positive free parameter. For simplicity, we take $\alpha$ in units of $(\mathrm{ns})^2$, $F$ in units of m/s, and $x$ and $y$ in units of nm; then the curvature image $L(x,y)$ when calculated on the function $F(r) = V(r)$has units of s/m in this work. In the limit of $\alpha \to 0$, $L(x,y)$ approaches the Laplacian with a negative sign, and with $\alpha \to \infty$, $L(x,y)$ will diverge at the extrema of $F$, giving $L(x,y)$ a good choice to trace the extrema position. When tracing the supercurrent network, we mainly focus on the maximal velocity rather than the minimal velocity; therefore, we apply a cutoff $L(x,y) \geq 0$ when applying this negative second derivative method to the supercurrent velocity image $V(x,y)$. We also employ a two-dimension Gaussian kernel with standard deviation $\sigma_r$ to smooth the $V(x,y)$ before computing the curvature map $L(x,y)$. Supplementary Fig. 1 shows a series of $L(x,y)$ maps with different $\sigma_r$ and $\alpha$.

**G. Topographies of the $Bi_2Sr_2CaCu_2O_8$ samples for different hole doping**

Extended Data Fig. 6 shows the topography images of the $Bi_2Sr_2CaCu_2O_8$ in the same field of view as in Fig. 2, with hole doping from $p$=0.17 to 0.22, from which the one-dimensional super-modulation gradually disappears with increasing the doping level. At the hole doping $p$=0.22, the unidirectional super-modulation is almost invisible.

**H. Separate imaging of supercurrent speeds for filled/empty states**

In theory for the superconducting density of states, the Galilean-boosted $N(E)$ is symmetric for the positive and negative energies when linear background augmentation is unimplemented, as it was constrained by the Bogoliubov-DeGennes Hamiltonian. To have a deeper inspection of the supercurrent speed identified in the positive and negative energy regimes, we separately symmetrize $g(E)$ from the positive and negative energy regimes and perform the VMAP analysis independently. Here it is important to note that the random variations in $g(r, E > 0)$ and $g(r, E < 0)$ which are generated by finite SNR are completely uncorrelated. Effectively, these are two independent experiments.

The resulting supercurrent velocity $V^+$ and $V^-$ are responses to the positive energy $g(E) > 0$ and negative energy $g(E) < 0$, respectively. In this work, the experimental spectra are always collected in an energy-symmetric window $[-E, +E]$, thus we can directly evaluate $V^\pm$ from the same procedure of the MAIN, and refinement on the COGNATE with 2000 steps on each of the spectra.

Extended Data Fig. 7 shows the $V^+(r)$ and $V^-(r)$ in the same FOV as in Fig. 2 with the hole doping $p$=0.17, 0.20, and 0.22, from which the $V^+(r)$ and $V^-(r)$ are virtually identical. Although these are effectively two separate independent experiments, the $V^-(r)$ images derived by VMAP from $g(r, E < 0)$ are virtually identical to the $V^+(r)$ images derived from $g(r, E > 0)$ in all studies, providing additional strong evidence of the validity of VMAP.

**I. Estimation of the effective magnetic field and comparison to reports**

In the simplest scenario, when the electron pairs move to form a closed circle loop, the current $I$ is:

$$I = \frac{nqV}{2\pi r} \tag{M11}$$

where $q = 2e$ is the effective charge in the supercurrent case and $e$ is the electron charge, $n$ is the number of flowing pairs, $r$ is the radius of the circle, and $V$ is the current velocity. Because the supercurrent circuits are only approximately 0.1 nm wide, no Meissner effect exists. Thus, the loop-current induced magnetic field from the Biot-Savart law is:

$$B = \frac{\mu_0 I}{2r} \tag{M12}$$

Where $\mu_0$ is the vacuum magnetic permeability. In Extended Data Figs. 5g-i, we show the histograms of the supercurrent velocity of $V(V > 0)$, where the averaged supercurrent velocity is $\langle V \rangle$~2100 m/s throughout the hole doping $p$=0.17 to $p$=0.22. To estimate the radius $r$ of the supercurrent loop, we compute the autocorrelation of $L(r)$ and the results are shown in Extended Data Fig. 8, from which the sharp peaks that close to the image center indicate the value of the diameter ($2r$) of the loops. The corresponding average radius of the supercurrent loop is $\langle r \rangle = 1.35(\pm 0.43)$ nm. By fitting the line cuts in the autocorrelation of $L(r)$ with an exponential function

$$y = Ae^{-\frac{d}{d_L}} + c \tag{M13}$$

where $A$ and $c$ are the free parameters and $d_L$ is the exponential decay length. The decay length $d_L$ from the center of autocorrelation of $L(r)$ is computed, from which the averaged $d_L$=0.14 nm (Extended Data Fig. 8), giving the width of the supercurrent loops. Meanwhile, from the normal-state band structure Eqn. 2, we estimate the Fermi wavevector $k_F \sim 4$ nm$^{-1}$. The electron density in an uncorrelated two-dimension electron gas is:

$$n_e = \frac{k_F^2}{2\pi} \tag{M14}$$

On the loop with radius $r$ and loop width $d_L$, the corresponding electron population is $2\pi r d_L n_e$, resulting in the Cooper pair population within the loop width

$$n_L = \frac{k_F^2 r d_L}{2} \tag{M15}$$

Taking the values of $k_F$, $r$, and $d_L$ into Eqn. M15, the estimated number of flowing pairs on a supercurrent loop is $n_L \sim 1.5$, the corresponding supercurrent intensity $I = 1 \times 10^{-7}$ A, and the corresponding effective magnetic field $|B| \approx 0.1$ mT. This estimated effective magnetic field is expected and consistent with the $\mu SR$-reported value[2,3].

On the other hand, magnetic dipoles beneath the surface of the sample might lead to current loops due to Meissner screening. However, a static moment $m$ at depth $d$ below the $CuO_2$ plane produces an in-plane vector potential whose maximum is $A_{\max} \approx 0.38\, \mu_0 m/(4\pi d^2)$[41]. The London response of the condensate to it would be a supercurrent of speed $V \approx (e/m_e)A$[42]. For $m = 5\, \mu_B$ at $d = 0.5$ nm this gives $V \approx 1$ m/s, and even for a cluster with $m = 100\, \mu_B$ only $V \approx 25$ m/s. Hence, this magnetic impurity phenomenon could only produce effects that are two to three orders of magnitude weaker than what we report in this work.

### J. "Rosetta Stone" of $d$-wave spectra in presence of a supercurrent

For a uniform superfluid velocity $V$ the condensate acquires a center-of-mass momentum $p_s = 2mV$, which Galilean boosts the quasiparticle spectrum $E_k = \pm\sqrt{{\xi_k}^2 + {\Delta_k}^2}$ ($\xi_k$ is the normal-state band dispersion) by

$$E_G(k) = \boldsymbol{V_F}(\boldsymbol{k}) \cdot \boldsymbol{p_s} \tag{M16}$$

The local density of states is then obtained from the imaginary part of the current-carrying Green's function,

$$N(\omega) = -\left(\frac{1}{\pi}\right) \Sigma_k \, Im\left[\frac{\omega + E_G(k) + \xi_k}{(\omega + \epsilon_G(k) + i0^+)^2 - \xi_k^2 - \Delta_k^2}\right] \tag{M17}$$

with $\Delta_k = \Delta_0 cos(2\varphi)$ for a $d_{x^2-y^2}$ superconductor over a circular Fermi surface. If the superfluid velocity makes an angle $\theta$ with the crystal $x$-axis, Fermi surface $\varepsilon_G(\varphi) = V_F p_s cos(\varphi - \theta)$, so the four nodal quasiparticles are shifted unequally according to their projection along the superfluid velocity. Consequently, the zero-bias density of states, which involves an integral over all $k_F(\varphi)$ on the Fermi surface, acquires a finite contribution proportional to the Galilean boost at the nodes.

On a nearly cylindrical Fermi surface, the gap takes the form $\Delta_k = \Delta_0 cos(2\varphi)$, so the excitation energy along the Fermi surface is $E(\varphi) = |\Delta_0 cos(2\varphi)|$. The Bogoliubov dispersion then possesses four symmetry-related saddle points at the antinodes, $k_{AN} = (\pm k_F, 0)$ and $(0, \pm k_F)$, where $\xi_k = 0$ and $|\Delta_k| = \Delta_0$. These saddle points give rise to van Hove singularities in the quasiparticle density of states at $\omega = \pm\Delta_0$, producing the familiar BCS coherence peaks. A finite superfluid velocity lifts the degeneracy of the four antinodal saddle points through the Galilean boost $E_G(k) = V_F p_s$.

For superfluid velocity directed along an antinodal axis ($\theta = 0$), the Fermi velocities at the antinodes $(0, \pm k_F)$ are perpendicular to the flow and therefore experience no Galilean boost, while those at $(\pm k_F, 0)$ are shifted by $\pm V_F p_s$. Consequently, each coherence peak evolves into a characteristic three-peak structure consisting of an unsplit central peak at $\omega = \pm\Delta_0$, arising from the two unaffected antinodes, together with two side peaks shifted upward and downward by the Galilean boost energy (Extended Data Fig. 9).

In contrast, when the superfluid velocity is directed along a nodal axis ($\theta = \pi/4$), all four antinodes acquire Galilean boost of equal magnitude $V_F p_s/\sqrt{2}$, with two shifted upward and two downward, so the central peak disappears and only a pair of split coherence peaks remains (Extended Data Fig. 9). Thus, the evolution from a three-peak spectrum for antinodal flow to a two-peak spectrum for nodal flow provides a direct spectroscopic signature of the orientation of the superfluid velocity relative to the $d_{x^2-y^2}$ order parameter.

**K. Supercurrent loop trajectories for different hole densities**

Extended Data Fig. 10 shows the supercurrent loop trajectories, $L(r)$ across the whole FOV, with hole dopings from $p$=0.17 to 0.22. For each FOV, we identify examples of individual supercurrent loops identified by the VMAP ANN. Colored dots indicate different directions relative to the crystal axes using the same concept as Figs. 1f-i. and are vertically shifted to add clarity. The correspondence between Extended Data Fig. 10, Fig. 3 and Figs. 1f-i provides additional confirmation of supercurrent flow in a $d$-wave superconductor across different doping and FOVs.

**L. Evaluation of the superfluid current conservation**

The supercurrent loop trajectories sometimes form tri-junctions throughout all FOV's studied in this paper. At each of these junctions, Kirchhoff's law can be evaluated to estimate if the superfluid current is conserved. The function $LV(r)$ is calculated by masking the $V(r)$ with the $L(r) > 0.5$ s/m map to get a speed-resolved network image. This is shown in Fig. 4a for $p$=0.17.

Due to the network pattern of $LV(r)$, a tri-junction can be distorted by nearby junctions, so a well-defined junction is selected with each branch being un-interfered with by other crossing branches, as shown in Figs. 4b-h. A linecut is then taken across each of the 3 branches, a little distance away from the junction and the average speed along the lines are extracted. In all cases where there are well-defined junctions, the speed of 1 branch is equal to the sum of the other two branches satisfying Kirchhoff's Law.

Separately, to evaluate the superfluid current conservation globally, we assume that the superfluid density does not vary strongly locally and can be taken as a constant value in the minimal patch unit. Then, Kirchhoff's law in the superfluid current velocity version can be described as evaluating the value $\sum_j V(j)$ for each site $i$, where the sum of observed supercurrent speeds through a site $i$ is derived from its four neighbor sites $j$. VMAP determines the larger and smaller of the two orthogonal velocity components, $V_H(j)$ and $V_L(j)$, at each pixel $j$. Therefore, the supercurrent speed along the horizontal axis at pixels $j = 1, 3$ is one of four values $\pm V_H(j)$; $\pm V_L(j)$, while the supercurrent speed along the vertical axis at pixels $j = 2, 4$ is also one of four values $\pm V_H(j)$; $\pm V_L(j)$. This means that for each of the 4 sites $j$ surrounding pixel $i$ there are 4 possible supercurrents. Thus there are $4^4 = 256$ possible combinations of $\pm V_H(j)$, $\pm V_L(j)$ for all four sites $j$ surrounding site $i$. For each pixel $i$ we then evaluate all possible values of:

$$S(i) = V_R^x - V_L^x + V_D^y - V_U^y \tag{M18}$$

where $R$, $L$, $D$ and $U$ denote the right, left, lower and upper neighbors spanning $j$. At site $i$, the relevant components ($V_R^x, V_L^x$ for $R$ and $L$; $V_D^y, V_U^y$ for $D$ and $U$) at each neighbor $j$ then take any of the four measured values $\pm V_H(j)$, $\pm V_L(j)$. $S(r_i)$ is then defined as the minimum value of $|S(i)|$ over all 256 possible combinations. In practice, we implement a minimal boxcar averaging on the superfluid current map $V(r)$ to enhance the signal-to-noise ratio before computing $S(r)$. The results of $S(r)$ are shown in the Fig. 4 and Supplementary Fig. 2, where the mean value of $S(r)$ is between 15 m/s and 20 m/s throughout all $p$, the corresponding ratio $\bar{S}/\bar{V}$ is about 1% with and 2% in all dopings studied in this work, providing evidence for universal supercurrent conservation throughout.

**M. Cross-correlation among different parameters from VMAP**

In Supplementary Fig. 2, we show all parameters in $r$-space obtained from the VMAP associated with the $\Delta(r)$ and $V(r)$ in a typical FOV with hole doping $p$ =0.20, and the corresponding normalized cross-correlation coefficients $X$ among different parameters are shown in Supplementary Table 2. The normalized cross-correlation coefficient $X$ is:

$$X(M,N) = \frac{\int M(r)N(r)dr}{\sqrt{\int |M(r)|^2 dr \int |N(r)|^2 dr}} \quad \text{(M19)}$$

where $M(r)$ and $N(r)$ are the target images after the mean-value subtracted. When we measure the normalized cross correlation $X(M,N)$ between $M(r)$ and $N(r)$ (where $M(r)$ and $N(r)$ are the output from VMAP), it is typically $X < 0.5$ for all samples studied (Supplementary Table 2).

**Methods-only References**

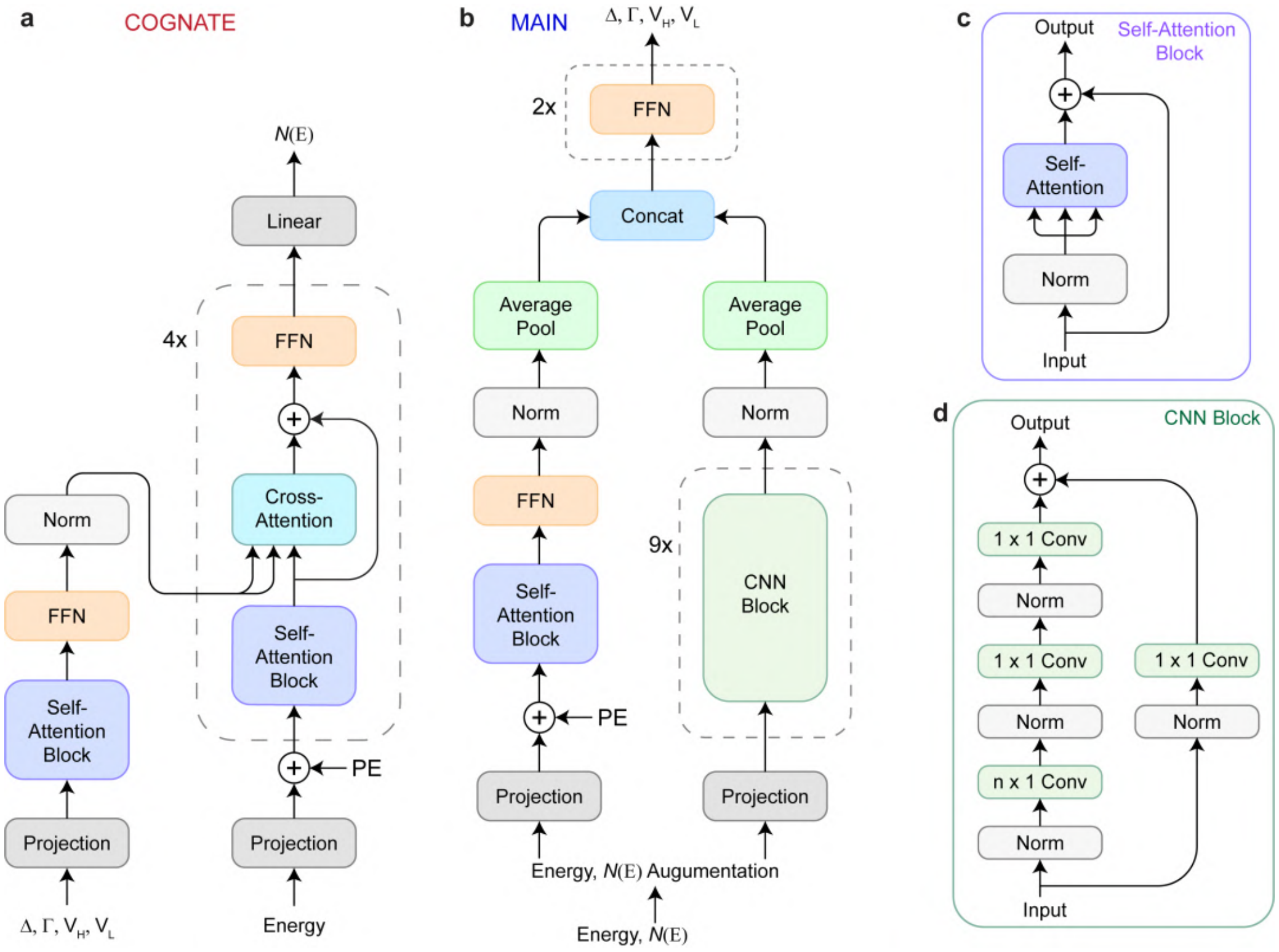

**Extended Data Fig. 1. VMAP architecture.**

(**a**) The architecture of the COGNATE, which takes the energy variable $E$, $\Delta$, $\Gamma$, $V_H$, and $V_L$ as input and outputs $N(E)$. (**b**) The architecture of the MAIN, which takes the energy and $N(E)$ as input, and outputs the four parameters: $\Delta$, $\Gamma$, $V_H$, and $V_L$. (**c**) The architecture of the Self-Attention block as in (**a**) and (**b**). (**d**) The architecture of Convolution-Neural-Network (CNN) block as in (**b**).

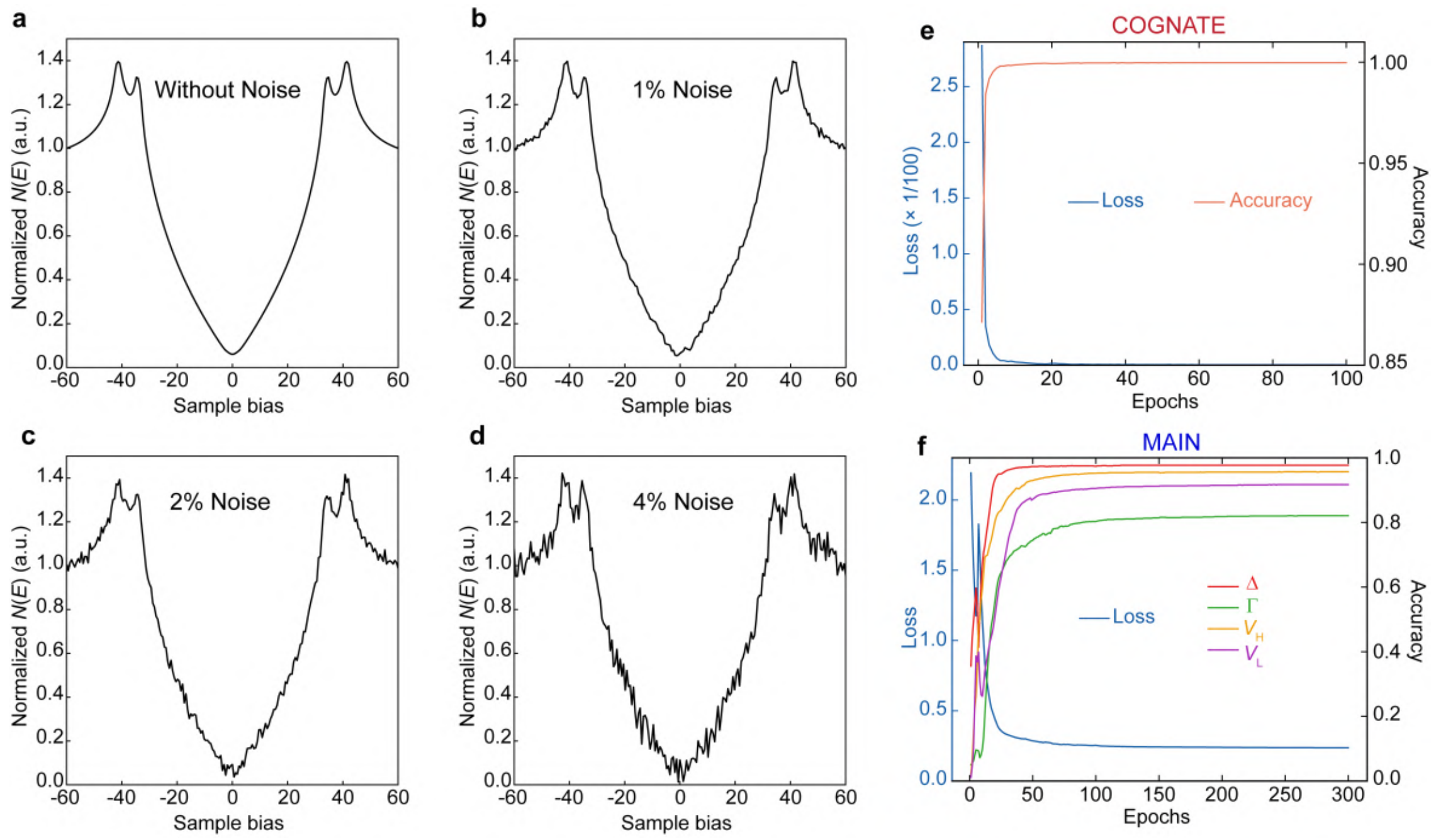

**Extended Data Fig. 2. Examples of data augmentation and the training of VMAP.**

(**a**) The example of clean spectrum without noise augmentation. (**b**-**d**) The examples of spectra with 1% (**b**), 2% (**c**), and 4% (**d**) noise augmentation. (**e**) The examples of validation loss and accuracy during training COGNATE. (**f**) The examples of validation loss and accuracy during training MAIN.

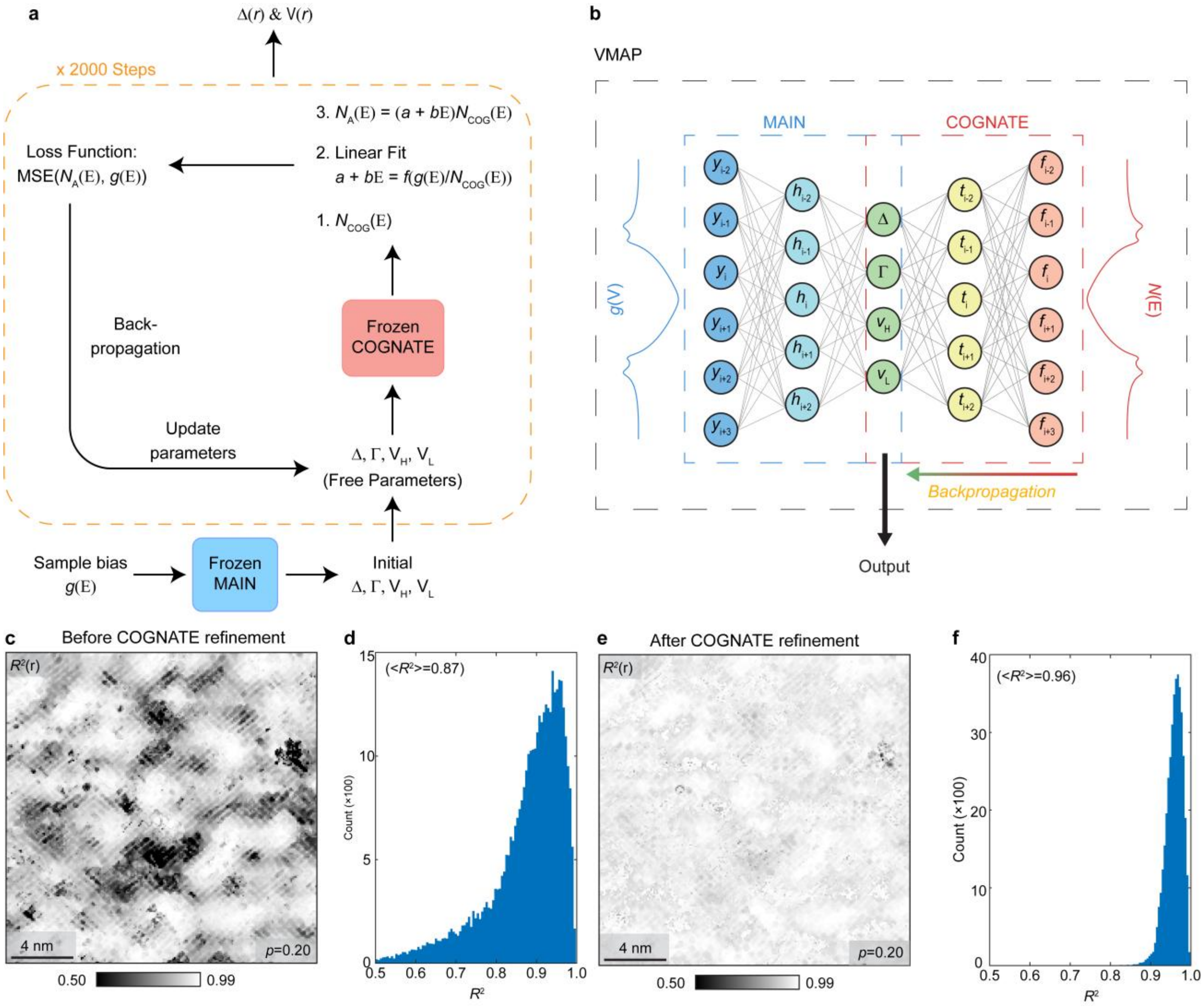

**Extended Data Fig. 3. The schematic workflow of VMAP and evaluation on Refinement**

(**a**) The schematic of VMAP refinement. The MAIN module derives the parameters from the experimental $g(r, E)$ spectra. Then those parameters, as the initial step, are input to the COGNATE model, where the background signal $a + bE$ is solved via linear fitting. The parameters are updated through the backpropagation mechanism. Each experimental spectrum is evaluated over 2000 refinement steps. (**b**) The schematic diagram of VMAP workflow. (**c**) A typical example of the spatial distribution of the coefficient of determination $R^2(r)$ without the COGNATE refinement. (**d**) The histogram of $R^2(r)$ as in (**c**), showing the mean value $\langle R^2 \rangle$=0.87. (**e**) The spatial distribution of the coefficient of determination $R^2(r)$ after the COGNATE refinement in the same FOV as (**c**). (**f**) The histogram of $R^2(r)$ as in (**e**), showing the mean value $\langle R^2 \rangle$=0.96.

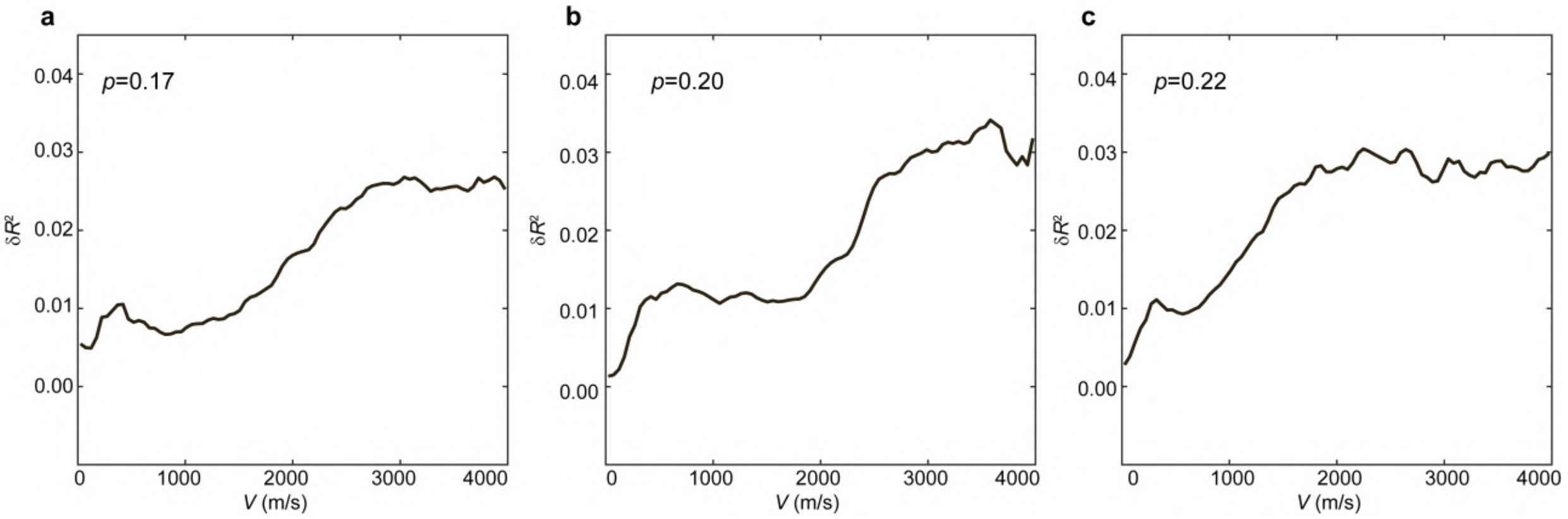

**Extended Data Fig. 4. Indistinguishable small supercurrents $V$<1250 m/s from $V$=0**

(**a**) The supercurrent velocity-dependent $\delta R^2 = R^2(V) - R^2(V = 0)$ in the sample with hole doping $p$=0.17, where the positive correlation indicates that the Galilean-boosted $N(E)$ provides a better description than $N(E)$ without supercurrent velocity, especially in the high supercurrent regime. (**b-c**) The same as in (**a**) but in the sample with hole doping $p$=0.20 (**b**) and $p$=0.22 (**c**). A boxcar average, with supercurrent velocity of 250 m/s, was applied to (**a-c**).

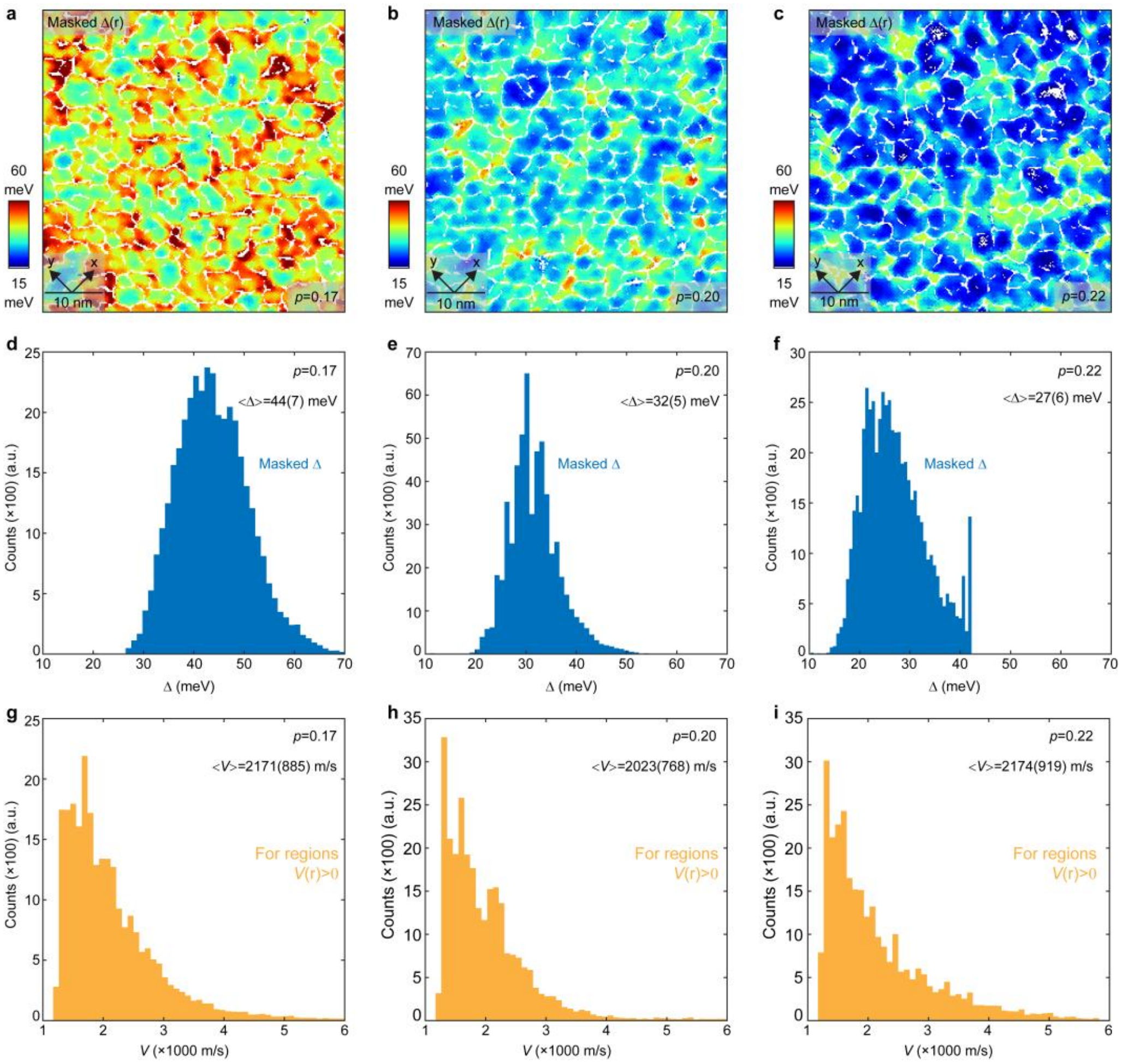

**Extended Data Fig. 5. Superconducting gap patterns and supercurrent velocities**

(**a-c**) The spatial distribution of the masked superconductive gaps $\Delta(r)$ with hole doping $p$=0.17 (**a**), $p$=0.20 (**b**), and $p$=0.22 (**c**), showing the patterns of the superconductive gap domains. (**d-e**) The histogram of superconductive gap from (**a-c**) with hole doping $p$=0.17 (**d**), $p$=0.20 (**e**) and $p$=0.22 (**f**), showing the averaged $\langle\Delta\rangle = 44(\pm 7)$ meV (**d**), $\langle\Delta\rangle = 32(\pm 5)$ meV (**e**), and $\langle\Delta\rangle = 27(\pm 6)$ meV (**f**), respectively. (**g-i**) The histogram of supercurrent velocity with hole doping $p$=0.17 (**g**), $p$=0.20 (**h**) and $p$=0.22 (**i**), showing the averaged $\langle V\rangle = 2171(\pm 885)$ m/s (**g**) $\langle V\rangle = 2023(\pm 768)$ m/s (**h**) and $\langle V\rangle = 2174(\pm 919)$ m/s (**i**), respectively.

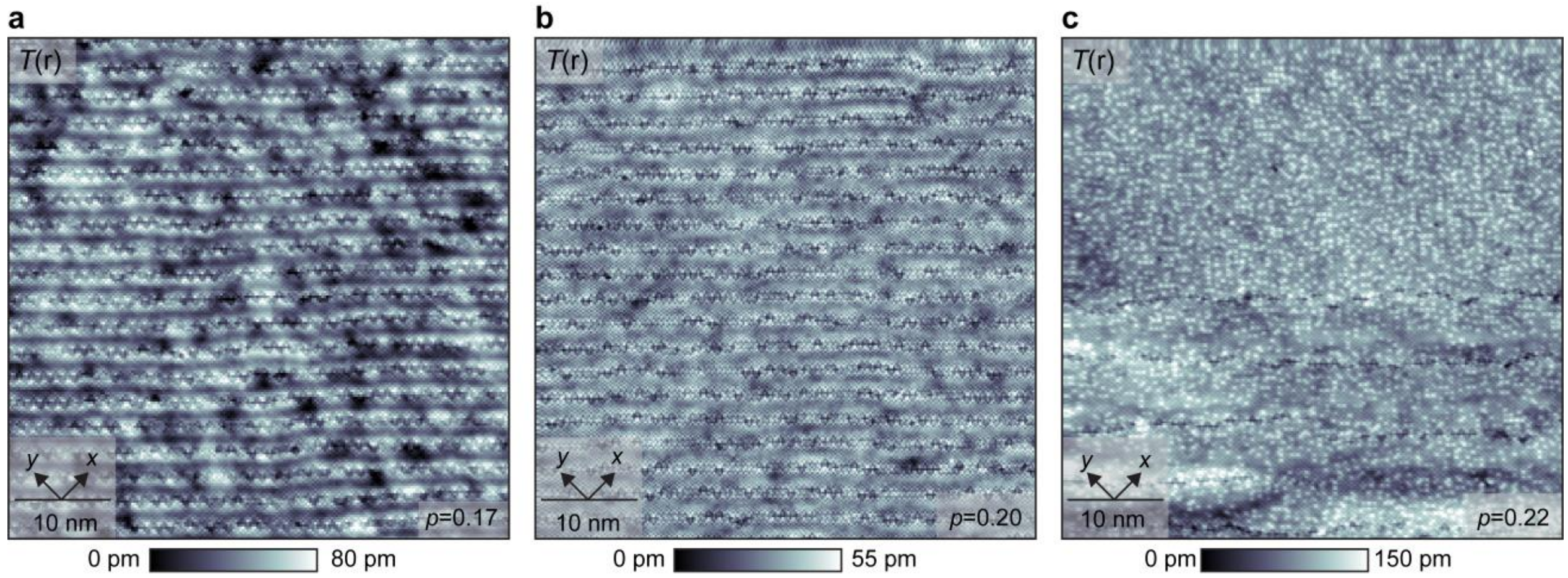

**Extended Data Fig. 6. Topography images of the $Bi_2Sr_2CaCu_2O_8$ with different hole doping**

(**a**) The topography image of optimal $Bi_2Sr_2CaCu_2O_8$ at doping $p$=0.17 in the same FOV as Figs. 2a, d, g ($V_{Bias}$=-90 mV, $I_t$=120 pA). (**b**) The topography image of over dopped $Bi_2Sr_2CaCu_2O_8$ at doping $p$=0.20 in the same FOV as Figs. 2b, e, h ($V_{Bias}$=60 mV, $I_t$=60 pA). (**c**) The topography image of over doped $Bi_2Sr_2CaCu_2O_8$ at doping $p$=0.22 in the same FOV as Figs. 2c, f, i ($V_{Bias}$=40 mV, $I_t$=60 pA).

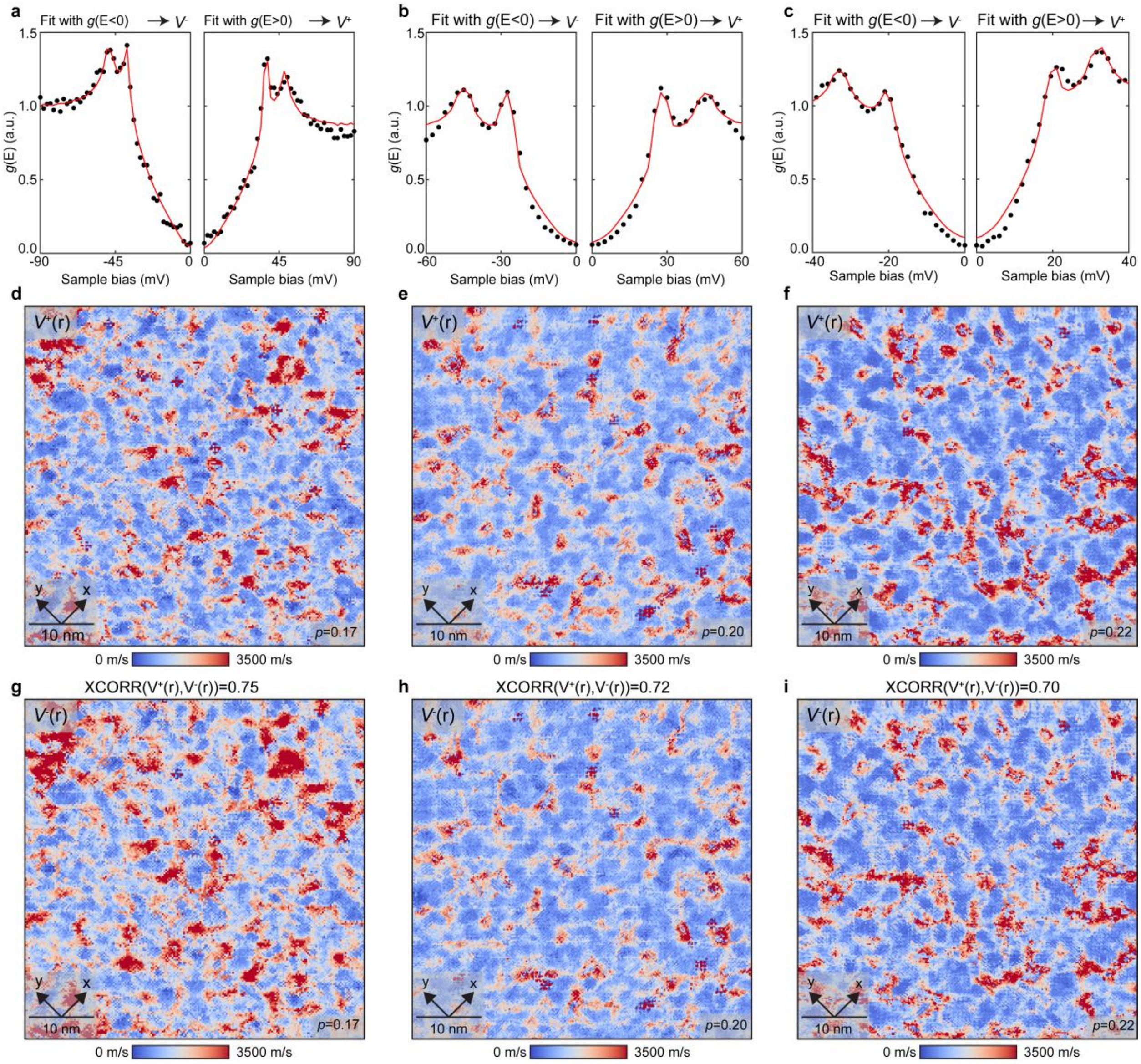

**Extended Data Fig. 7. $\boldsymbol{V^+(r)}$ and $\boldsymbol{V^-(r)}$ identified by the VMAP**

(**a–c**) Example spectra at $p$ =0.17 (**a**), $p$=0.20 (**b**) and $p$=0.22 (**c**), fitted separately for $E < 0$ (giving $V^-$) and for $E > 0$ (giving $V^+$). (**d–f**) The $V^+(r)$ identified by the VMAP with different hole doping $p$=0.17 (**d**), $p$=0.20 (**e**), and $p$=0.22 (**f**). (**g-i**) The $V^-(r)$ identified by the VMAP with different hole doping $p$=0.17 (**g**), $p$=0.20 (**h**), and $p$=0.22 (**i**) in the same FOV as in (**d-f**).

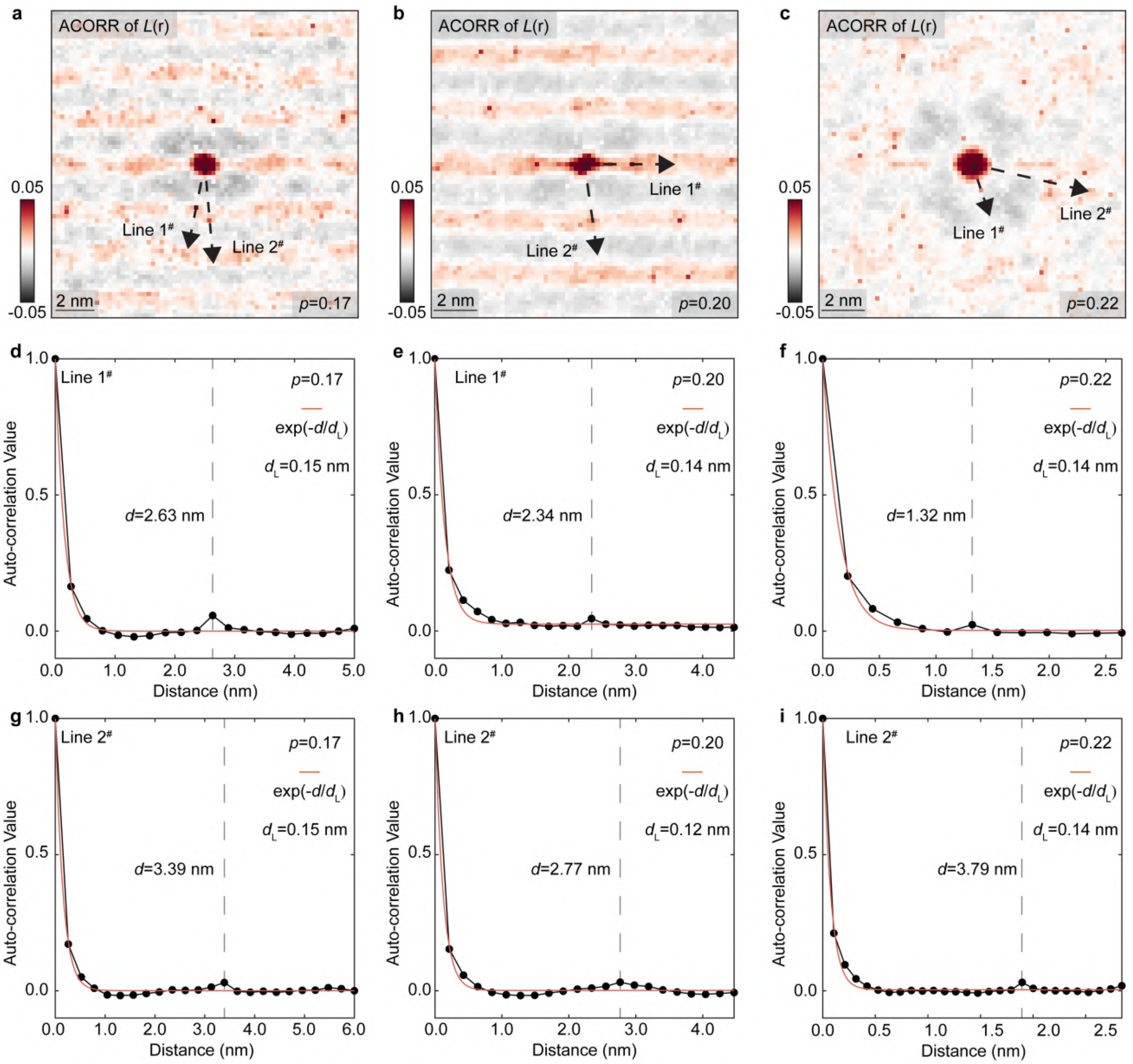

**Extended Data Fig. 8. Estimation of the supercurrent loop radius.**

(**a**) The auto-correlation map of $L(r)$ with hole doping $p$=0.17. (**b**) The auto-correlation map of $L(r)$ with hole doping $p$=0.20. (**c**) The auto-correlation map of $L(r)$ with hole doping $p$=0.22. (**d-i**) The line cuts of $L(r)$ marked in (**a-c**), indicating the averaged supercurrent loop radius $r = 1.4$ nm (half of the dimeter), and the exponential function fitting (the solid orange curves) give the averaged loop with $d_L$=0.14 nm.

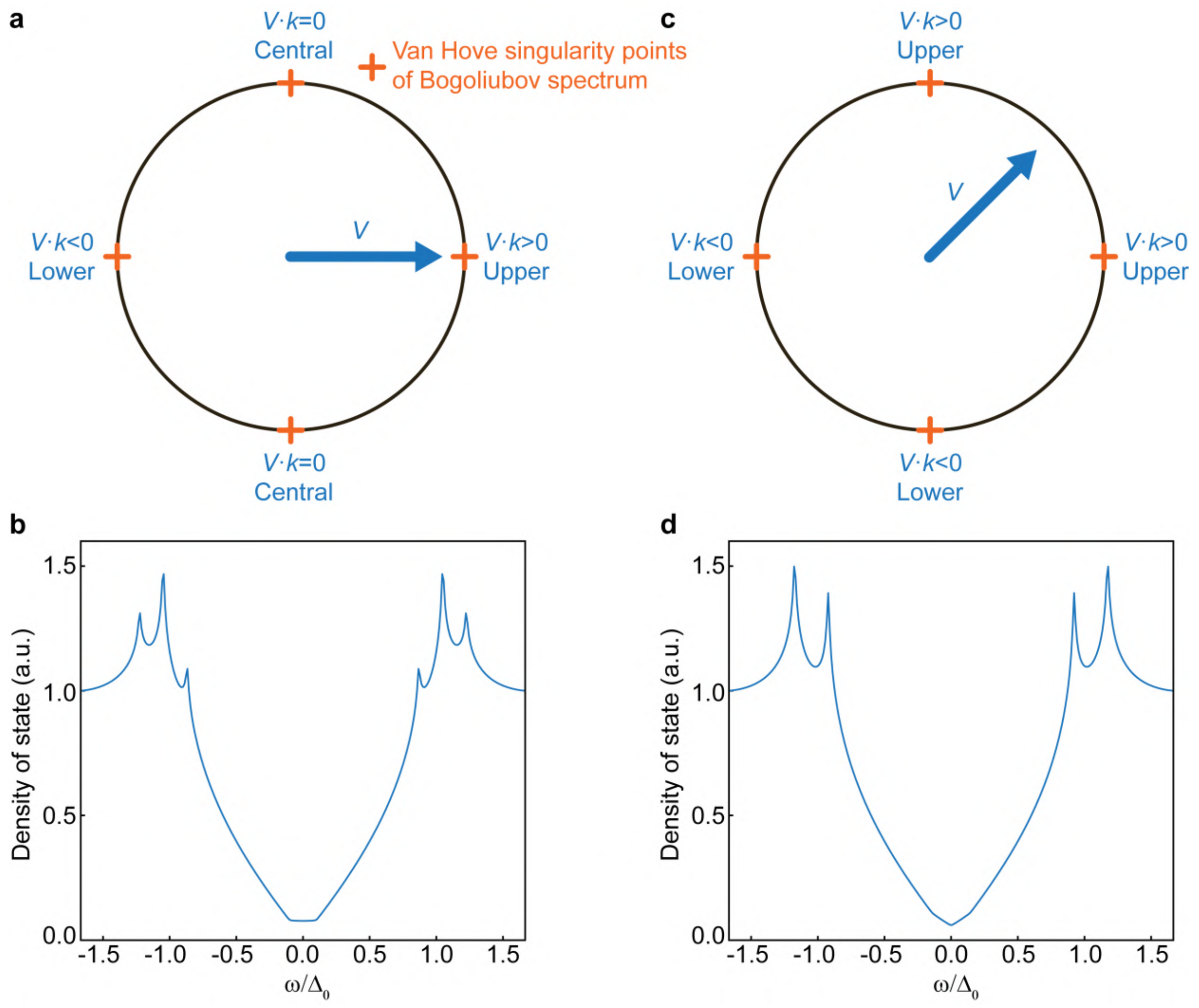

**Extended Data Fig. 9 Dependence of local spectra on superfluid velocity**.

(**a-b**) Velocity along antinode (**a**) and corresponding spectrum (**b**). (**c-d**) Velocity along node (**c**) and corresponding spectrum (**d**).

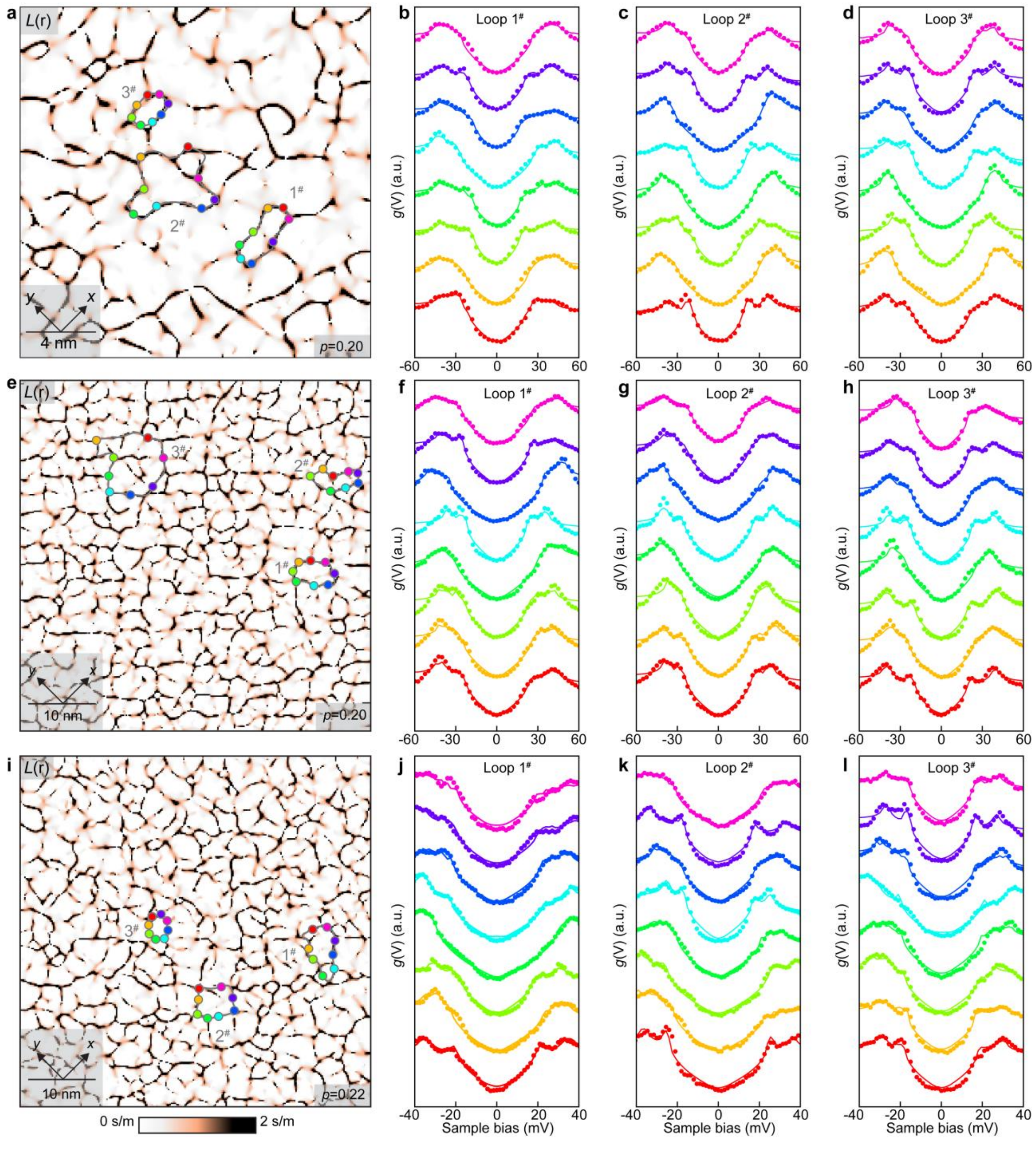

**Extended Data Fig. 10. Examples of individual supercurrent loop spectra**

(**a**) The $L(r)$ as in the Supplementary Figs. 1-2, where the hole doping $p$=0.20. (**b-d**) The examples of individual supercurrent loops identified by the VMAP, the loops $1^{\#}$-$3^{\#}$ are marked in (**a**) ($V_{Bias}$=60 mV, $I_t$=60 pA, $V_{mod}$=3.0 mV). The spectra in (**b-d**) are vertically shifted for clarity. (**e**) The $L(r)$ as in the Fig. 2, where the hole doping $p$=0.20. (**f-h**) The examples of individual supercurrent loops identified by the VMAP, the loops $1^{\#}$-$3^{\#}$ are marked in (**e**) ($V_{Bias}$=60 mV, $I_t$=60 pA, $V_{mod}$=2.5 mV). The spectra in (**f-h**) are vertically shifted for clarity. (**i**) The $L(r)$ as in the Fig. 2, where the hole doping $p$=0.22. (**j-l**) The examples of individual supercurrent loops identified by the VMAP, the loops $1^{\#}$-$3^{\#}$ are marked in (**i**) ($V_{Bias}$=40 mV, $I_t$=60 pA, $V_{mod}$=2.5 mV). The spectra in (**j-l**) are vertically shifted for clarity.

**Supplementary Information**

**Table 1**

| $p$ | $\mu\ (meV)$ | $t_1\ (meV)$ | $t_2\ (meV)$ | $t_3\ (meV)$ |
|---|---|---|---|---|
| 0.17 | 407 | 360 | 108 | 36 |
| 0.20 | 414 | 360 | 108 | 36 |
| 0.22 | 427 | 360 | 108 | 36 |

**Supplementary Table 1.**

The tight binding parameters for $Bi_2Sr_2CaCu_2O_8$ with hole doping $p$=0.17, 0.20, and 0.22 studied in this work.

**Table 2**

| $\boldsymbol{X}$ | $\boldsymbol{p = 0.17}$ | $\boldsymbol{p = 0.20}$ | $\boldsymbol{p = 0.22}$ | $\boldsymbol{p = 0.20}$ **(Small FOV)** |
|---|---|---|---|---|
| $\Delta - V$ | 0.49 | 0.44 | 0.56 | 0.40 |
| $\Delta - \Gamma$ | 0.33 | 0.25 | 0.28 | 0.39 |
| $\Delta - a$ | 0.55 | 0.04 | -0.17 | -0.06 |
| $\Delta - b$ | 0.55 | 0.52 | 0.35 | 0.50 |
| $V - \Gamma$ | 0.28 | 0.09 | 0.19 | 0.13 |
| $V - a$ | 0.20 | -0.14 | -0.40 | -0.26 |
| $V - b$ | 0.33 | 0.35 | 0.31 | 0.33 |
| $\Gamma - a$ | 0.24 | -0.01 | -0.22 | 0.02 |
| $\Gamma - b$ | 0.21 | 0.16 | 0.10 | 0.24 |
| $a - b$ | 0.77 | -0.52 | -0.54 | -0.52 |

**Supplementary Table 2.**

The cross-coefficient among different parameters obtained from VMAP for different hole doping $p$.

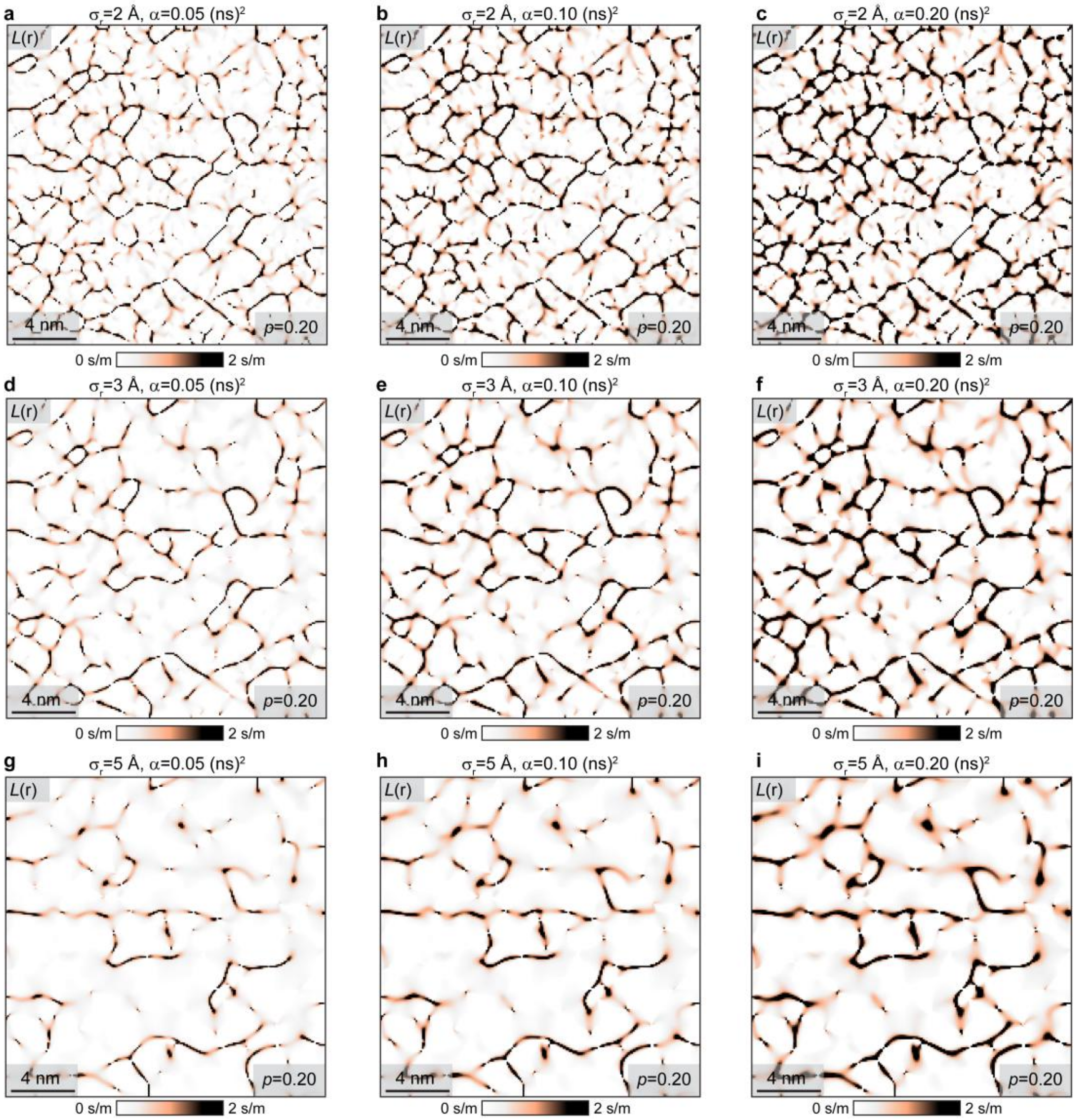

**Supplementary Fig. 1. $L(r)$ maps at different $\sigma_r$ and $\alpha$**

(**a**-**i**) The $L(r)$ maps for different $\sigma_r$ from 2.0 Å to 5.0 Å and at different $\alpha$ from 0.05(ns)$^2$ to 0.20 (ns)$^2$.

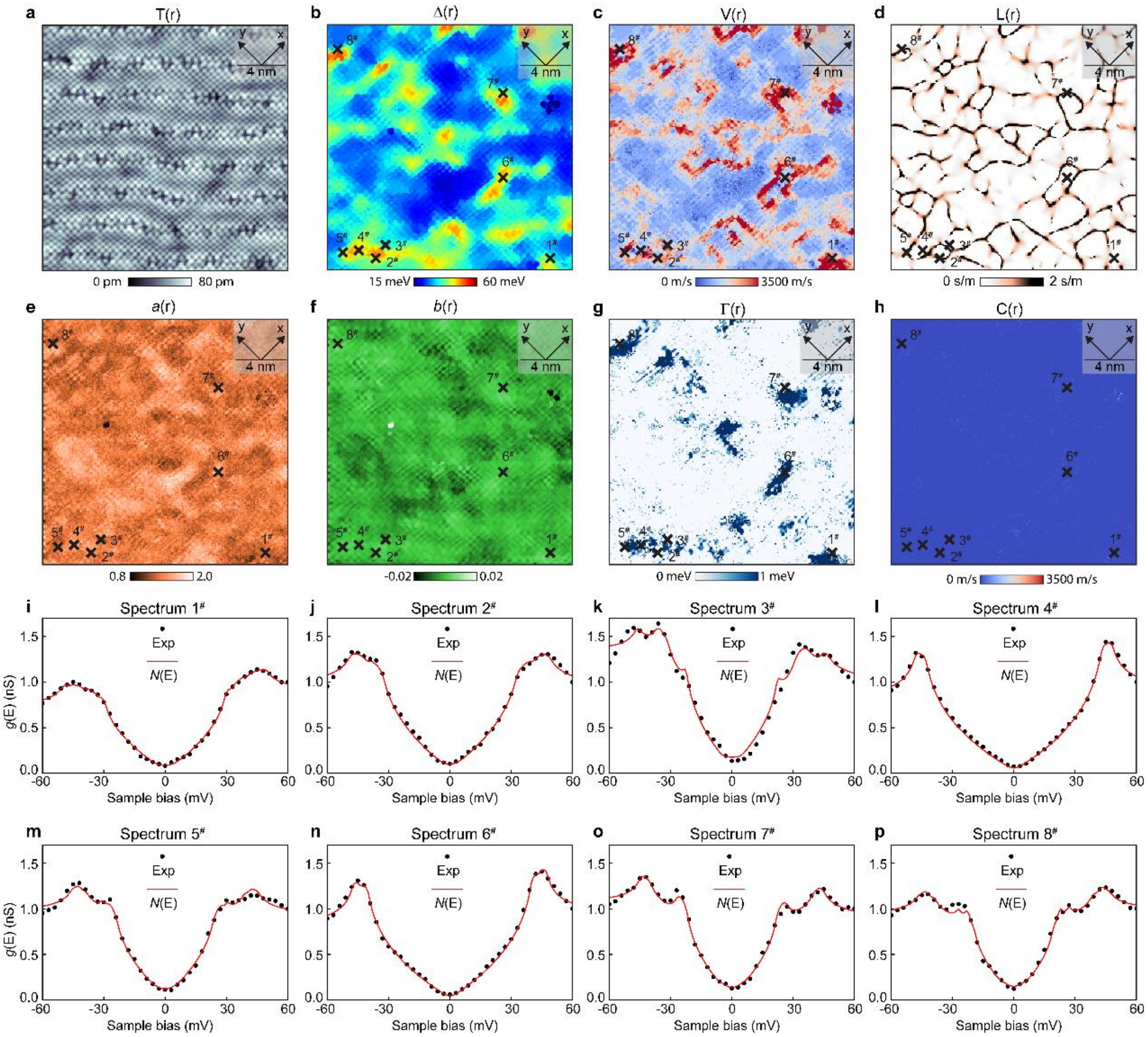

**Supplementary Fig. 2 Simultaneous determination of different parameters in a typical FOV**

The VMAP analysis with hole doping $p$=0.20 in the same FOV of Supplementary Fig. 1. (**a**) The topography image. (**b-h**) The parameters obtained from VMAP in the same field FOV as in (**a**). (**i-p**) A series of example spectra obtained from VMAP, where the positions are marked in (**b-h**) ($V_{Bias}$=60 mV, $I_t$=60 pA, $V_{mod}$=3 mV).

## Supplementary Videos

To demonstrate validity of determination of $\Delta(r)$ and $V(r)$ by the VMAP ANN we have prepared three-panel Video of simultaneous $g(r,E)$: $N(r,E)$: $|g(r,E) - N(r,E)|$ : Supplementary Video 1 for p=0.17, Supplementary Video 2 for p=0.20, and Supplementary Video 3 for p=0.22. Because the $|g(r,E) - N(r,E)|$ manifestly yields values that are virtually zero for all $E$, this demonstrates directly and visually the excellent quality of the fits to $g(r,E)$, which are generated by VMAP assignment of $N(r,E)$, throughout this work.